\def\llncs{0}
\def\fullpage{1}
\def\anonymous{0}
\def\authnote{0}
\def\draft{0} 
\def\notxfont{0}
\def\submission{0} 

\ifnum\submission=1
\def\anonymous{1}
\def\llncs{1}
\def\fullpage{0}
\fi

\ifnum\llncs=1 	
    \documentclass[envcountsect,a4paper,runningheads]{llncs}
\else
    \documentclass[letterpaper,hmargin=1.05in,vmargin=1.05in,11pt]{article}
    \ifnum\fullpage=1
        \usepackage{fullpage}
    \fi
\fi

\usepackage[%
  colorlinks=true,
  citecolor=blue,
  backref=page
]{hyperref}

\usepackage{amsmath, amsfonts, amssymb, mathtools,amscd}

\usepackage{amsthm}

\usepackage{lmodern}
\usepackage[T1]{fontenc}
\usepackage[utf8]{inputenc}

\usepackage{arydshln} 
\usepackage{url}
\usepackage{ifthen}
\usepackage{bm}
\usepackage{multirow}
\usepackage[dvips]{graphicx}
\usepackage[usenames]{color}
\usepackage{xcolor,colortbl} 
\usepackage{threeparttable}
\usepackage{comment}
\usepackage{paralist,verbatim}
\usepackage{cases}
\usepackage{booktabs}
\usepackage{breakcites}
\usepackage{braket}
\usepackage{cancel} 
\usepackage{ascmac} 
\usepackage{framed}
\usepackage{authblk}
\usepackage{pifont}
\usepackage{qcircuit}
\usepackage{tikz}
\usepackage{xspace} 
\usetikzlibrary{decorations.pathmorphing,decorations.shapes,decorations.markings}
\definecolor{darkblue}{rgb}{0,0,0.6}
\definecolor{darkgreen}{rgb}{0,0.5,0}
\definecolor{maroon}{rgb}{0.5,0.1,0.1}
\definecolor{dpurple}{rgb}{0.2,0,0.65}

\usepackage[capitalise,noabbrev]{cleveref}
\usepackage[absolute]{textpos}
\usepackage[final]{microtype}
\usepackage[absolute]{textpos}
\usepackage{everypage}
\DeclareMathAlphabet{\mathpzc}{OT1}{pzc}{m}{it}

\usepackage{algorithmic}
\usepackage{algorithm}
\usepackage{here}

\ifnum\draft=1
    \usepackage[color]{showkeys}
    \definecolor{refkey}{cmyk}{0,0,0,.25}
    \definecolor{labelkey}{cmyk}{0,0,0,.7}
    \usepackage{CJKutf8}
\else
\fi

\newtheoremstyle{thicktheorem}%
{\topsep}
{\topsep}
{\itshape}{}%
{\bfseries}%
{.}
{ }%
{\thmname{#1}\thmnumber{ #2}%
		\thmnote{ (#3)}%
}

\newtheoremstyle{remark}
{\topsep}
{\topsep}
	{}
	{}
	{}
	{.}
	{ }
	{\textit{\thmname{#1}}\thmnumber{ #2}
			\thmnote{ (#3)}%
	}

\usepackage{aliascnt}
\ifnum\llncs=0
	\theoremstyle{thicktheorem}
	\newtheorem{theorem}{Theorem}[section]
	\newaliascnt{lemma}{theorem}
	\newtheorem{lemma}[lemma]{Lemma}
	\aliascntresetthe{lemma}
	\newaliascnt{corollary}{theorem}
	\newtheorem{corollary}[corollary]{Corollary}
	\aliascntresetthe{corollary}
	\newaliascnt{proposition}{theorem}
	
	\aliascntresetthe{proposition}
	\newaliascnt{definition}{theorem}
	\newtheorem{definition}[definition]{Definition}
	\aliascntresetthe{definition}
	\newaliascnt{game}{theorem}
	
	\aliascntresetthe{game}

	\theoremstyle{remark}
	\newaliascnt{claim}{theorem}
	\newtheorem{claim}[claim]{Claim}
	\aliascntresetthe{claim}
	\newaliascnt{remark}{theorem}
	\newtheorem{remark}[remark]{Remark}
	\aliascntresetthe{remark}
	\newaliascnt{example}{theorem}
	
	\aliascntresetthe{example}
	\newaliascnt{note}{theorem}
	
	\aliascntresetthe{note}
	\newaliascnt{conjecture}{theorem}
	
	\aliascntresetthe{conjecture}
\else
\fi
\newaliascnt{MyClaim}{theorem}

\aliascntresetthe{MyClaim}
\Crefname{MyClaim}{Claim}{Claims}

	\crefname{theorem}{Theorem}{Theorems}
	\crefname{assumption}{Assumption}{Assumptions}
	\crefname{construction}{Construction}{Constructions}
	\crefname{corollary}{Corollary}{Corollaries}
	\crefname{conjecture}{Conjecture}{Conjectures}
	\crefname{definition}{Definition}{Definitions}
	\crefname{exmaple}{Example}{Examples}
	\crefname{experiment}{Experiment}{Experiments}
	\crefname{counterexample}{Counterexample}{Counterexamples}
	\crefname{lemma}{Lemma}{Lemmata}
	\crefname{observation}{Observation}{Observations}
	\crefname{proposition}{Proposition}{Propositions}
	\crefname{remark}{Remark}{Remarks}
	\crefname{claim}{Claim}{Claims}
	\crefname{fact}{Fact}{Facts}
	\crefname{note}{Note}{Notes}

\ifnum\llncs=1
 \crefname{appendix}{App.}{Appendices}
 \crefname{section}{Sec.}{Sections}
\else
\fi

\ifnum\llncs=1
\renewcommand*{\backref}[1]{}
\else
	\renewcommand*{\backref}[1]{(Cited on page~#1.)}
	\ifnum\notxfont=1
	\else
		\usepackage{newtxtext}
	\fi
\fi
\ifnum\authnote=0  
\newcommand{\mor}[1]{}
\newcommand{\minki}[1]{}
\newcommand{\takashi}[1]{}
\newcommand{\xagawa}[1]{}

\else
\newcommand{\mor}[1]{$\ll$\textsf{\color{red} Tomoyuki: { #1}}$\gg$}
\newcommand{\takashi}[1]{$\ll$\textsf{\color{orange} Takashi: { #1}}$\gg$}

\newcommand{\xagawa}[1]{$\ll$\textsf{\color{magenta} Keita: { #1}}$\gg$}

\fi

\newcommand{\Tr}{\mathrm{Tr}}

\newcommand{\sigk}{\mathsf{sigk}}

\newcommand{\Samp}{\algo{Samp}}

\newcommand{\norm}[1]{\left\|#1\right\|}

\newcommand{\cA}{\mathcal{A}}

\def\makeuppercase#1{
\expandafter\newcommand\csname tl#1\endcsname{\widetilde{#1}}
}

\def\makelowercase#1{
\expandafter\newcommand\csname tl#1\endcsname{\widetilde{#1}}
}

\newcommand{\regC}{\mathbf{C}}

\newcommand{\regM}{\mathbf{M}}
\newcommand{\regB}{\mathbf{B}}
\newcommand{\regA}{\mathbf{A}}
\newcommand{\regE}{\mathbf{E}}

\newcommand{\regS}{\mathbf{S}}
\newcommand{\regW}{\mathbf{W}}

\newcommand{\secp}{\lambda}

\newcommand*{\vk}{\keys{vk}}

\newcommand*{\keys}[1]{\mathsf{#1}}

\newcommand*{\algo}[1]{\ensuremath{\mathsf{#1}}}

\newenvironment{boxfig}[2]{\begin{figure}[#1]\fbox{\begin{minipage}{0.97\linewidth}
                        \vspace{0.2em}
                        \makebox[0.025\linewidth]{}
                        \begin{minipage}{0.95\linewidth}
            {{
                        #2 }}
                        \end{minipage}
                        \vspace{0.2em}
                        \end{minipage}}}{\end{figure}}

\newcommand{\bit}{\{0,1\}}

\newcommand{\KeyGen}{\algo{KeyGen}}

\newcommand{\Sign}{\algo{Sign}}
\newcommand{\Vrfy}{\algo{Vrfy}}

\newcommand{\Ver}{\algo{Ver}}

\newcommand{\negl}{{\mathsf{negl}}}

\newcommand{\poly}{{\mathrm{poly}}}

\makeatletter
\DeclareRobustCommand
  \myvdots{\vbox{\baselineskip4\p@ \lineskiplimit\z@
    \hbox{.}\hbox{.}\hbox{.}}}
\makeatother
\usetikzlibrary{positioning, arrows.meta, calc, decorations.markings}

\title{Quantum Key Exchange from Public Quantum Money}

\ifnum\anonymous=1
\ifnum\llncs=1
\author{\empty}\institute{\empty}
\else
\author{}
\fi
\else
\ifnum\llncs=1
\author{
Tomoyuki Morimae\inst{1}
} 
\institute{
 Yukawa Institute for Theoretical Physics, Kyoto University, Kyoto, Japan 
}
\else
\author[1]{Eli Goldin}
\author[2]{Tomoyuki Morimae}
\affil[1]{{\small Princeton University, USA}\authorcr{\small
eg0439@princeton.edu
} }
\affil[2]{{\small Kyoto University, Japan}\authorcr{\small tomoyuki.morimae@yukawa.kyoto-u.ac.jp} }
\fi 
\fi

\date{}

\begin{document}

\maketitle

\begin{abstract}
Two-message quantum key exchange (2QKE) cannot be realized in a statistically secure way, 
and therefore some computational assumptions are required [Malavolta and Walter 2024]. 
So far, the known weakest assumption for 2QKE is the existence of one-way functions (OWFs): 
Malavolta and Walter constructed (even everlasting-secure) 2QKE from OWFs [Malavolta and Walter 2024],
and Kitagawa, Morimae, Nishimaki, and Yamakawa constructed quantum public-key encryption (QPKE) with quantum public keys,
which implies 2QKE, from OWFs [Kitagawa, Morimae, Nishimaki and Yamakawa 2024].
Can we construct 2QKE from a weaker assumption than OWFs?

In this paper, we give the first complete characterization of 2QKE.
We show that the following four primitives are (existentially) equivalent:
everlasting-secure 2QKE, standard 2QKE, efficiently-verifiable state puzzles (EV-StatePuzzs),
and one-time secure digital signatures with quantum signatures and signing keys.

Because EV-StatePuzzs are potentially weaker than OWFs [Kretschmer, Qian and Tal 2025], 
the equivalence implies that 2QKE are potentially weaker than OWFs.
Moreover, our result shows the equivalence between standard 2QKE and everlasting-secure 2QKE, which
means that everlasting security comes for free.
Finally, one-shot signatures, quantum lightning, and public-key quantum money mini-schemes imply EV-StatePuzzs, 
and 
therefore, 
our result also gives a new lowerbound, namely everlasting-secure 2QKE, of these public-key unclonable primitives.

We also show that one-shot signatures imply QPKE with quantum public keys.
[Kitagawa, Morimae, Nishimaki and Yamakawa, 2024]
constructed QPKE with quantum public keys from OWFs, and it was open whether
QPKE with quantum public keys can be constructed from an assumption weaker than or incomparable to OWFs.
Our result gives a potential answer to this question since
one-shot signatures seem to be incomparable with OWFs.
Moreover, QPKE constructed from one-shot signatures has an additional functionality, which we call certified freeze,
that allows the receiver to publish 
a publicly verifiable classical certificate that computationally binds the receiver 
to an upper bound on the total number of distinct valid ciphertexts under its verification key.
\end{abstract}

\ifnum\submission=0
\newpage
\setcounter{tocdepth}{2}
\tableofcontents
\newpage
\fi

\section{Introduction}


\paragraph{Key exchange.} Key exchange (KE), introduced in the seminal work of Diffie and
Hellman~\cite{DifHel76}, allows two parties, Alice and Bob, who share no secret
information in advance, to agree on a common secret key by communicating over
public authenticated channels. It is one of the most fundamental primitives in cryptography.
In the
classical world, however, KE is a ``Cryptomania'' object in Impagliazzo's
terminology~\cite{Imp95}: all known constructions rely on structured,
public-key-type hardness assumptions, such as the hardness of the discrete
logarithm problem~\cite{DifHel76}, factoring, and learning with
errors~\cite{STOC:Regev05}. This is believed to
be inherent: Impagliazzo and Rudich~\cite{STOC:ImpRud89} showed that KE cannot
be constructed from one-way functions (OWFs) in a black-box way (see
also~\cite{C:BarMah09}), and hence classical KE seems out of reach of
Minicrypt assumptions.

The situation changes drastically once quantum communication is available. The
celebrated BB84 protocol~\cite{BB84} allows Alice and Bob to agree on a shared
key with information-theoretic security, assuming only authenticated classical
channels; the quantum channel may be completely unauthenticated.\footnote{Quantum authentication implies security, and therefore
if a quantum authenticated channel is assumed, a secure communication is trivially achieved. Hence, for quantum PKE and quantum KE, 
an authenticated quantum channel should not be assumed.} This, however,
comes at the cost of interaction: BB84 and its descendants require more than
two messages, whereas classical KE can be achieved with only two messages,
which is optimal. 
In fact,
no two-message quantum key exchange (2QKE) protocol---where Alice sends a single (quantum and classical) message to Bob
over an unauthenticated quantum channel and an authenticated classical channel,
and Bob replies with a single message---can be secure against unbounded
adversaries~\cite{C:MalWal24}. (See also \cite{EC:QiaRaiZha25,cryptoeprint:2024/1707}.)
Computational assumptions are therefore necessary in the
two-message setting, and it is natural to ask which computational assumptions
suffice.

Malavolta and Walter~\cite{C:MalWal24} constructed a 2QKE (in fact, with everlasting security) assuming only 
OWFs, and Kitagawa, Morimae, Nishimaki, and Yamakawa~\cite{C:KMNY24}
constructed quantum public-key encryption with quantum public
keys from OWFs, which also yields 2QKE in the same setting. 
In summary, in the
standard setting where only classical channels are authenticated, OWFs are
currently the weakest assumption known to imply 2QKE. On the other hand, one of
the most remarkable lessons of quantum cryptography in recent years is that
OWFs need not be the minimal assumption~\cite{Kre21,C:MorYam22,C:AnaQiaYue22}: many primitives, including
secret-key encryptions, commitments, oblivious transfer, and multi-party computation, can be based on
assumptions that are potentially weaker than OWFs, such as pseudorandom
states~\cite{C:JiLiuSon18}, one-way
puzzles~\cite{STOC:KhuTom24}, and EFI pairs~\cite{ITCS:BCQ23}, and many of these
primitives can exist even if OWFs do not~\cite{Kre21,STOC:KQST23,STOC:KreQiaTal25,STOC:LomMaWri24}. However, no such ``Microcrypt'' primitive is
known to imply 2QKE, and therefore the following question
has remained open.
\begin{center}
\emph{Can we construct 2QKE from an assumption weaker than OWFs?} 
\end{center}

\paragraph{Public-key encryption.}
Public-key encryption (PKE)~\cite{DifHel76,RivShaAdl78,STOC:GolMic82} is a strengthening of KE:
Alice publishes a single public key, and anyone holding it can encrypt a message that only Alice can decrypt.
Thus, whereas KE lets two designated parties agree on a secret, PKE lets Alice receive secret messages from
arbitrarily many senders without any prior contact with them.
In the classical world, PKE is again a Cryptomania object~\cite{Imp95}: PKE trivially implies KE, and hence
the black-box separation of Impagliazzo and Rudich~\cite{STOC:ImpRud89} rules out constructions of PKE from
OWFs as well.
Once quantum communication is available, however, the picture changes as it does for KE.
Kitagawa, Morimae, Nishimaki, and Yamakawa~\cite{C:KMNY24} showed that, if public keys are allowed to be
quantum states, quantum PKE (QPKE) can be constructed from OWFs alone, even when unbounded polynomially many copies
of quantum public keys are transmitted
over unauthenticated quantum channels.
In other words, quantum physics can move PKE from Cryptomania down to Minicrypt.
It is then natural to ask whether it can descend even further, into Microcrypt, and hence, in parallel
with 2QKE, we have the following open problem.
\begin{center}
\emph{Can we construct QPKE with quantum public keys from an assumption weaker than or incomparable to OWFs?} 
\end{center}

\paragraph{Public-key quantum money and its variants.}
Quantum physics 
enables entirely new primitives that are classically impossible,
based on the no-cloning theorem.
Public-key quantum money~\cite{STOC:AarChr12} is a scheme where quantum states,
called banknotes, are issued together with classical serial numbers:
anyone can publicly verify a banknote with its serial number, but no one can duplicate it.
Aaronson and Christiano~\cite{STOC:AarChr12} abstracted its core as a \emph{mini-scheme},
which consists of a minting algorithm that outputs a pair of a classical serial number and a banknote,
and a verification algorithm;
the security requires that no QPT adversary, given a single banknote,
can output a (possibly entangled) state that passes the verification
on both of two registers with respect to the same serial number.
Quantum lightning~\cite{JC:Zhandry21} is a strengthening of public-key quantum money mini-schemes:
no QPT adversary can output two states that pass the verification
with respect to the same serial number,
even though the adversary prepares the states by itself.
One-shot signatures (OSS)~\cite{STOC:AGKZ20} further strengthen quantum lightning
into a signature-type primitive:
given a classical common reference string,
anyone can generate a classical verification key together with a quantum signing key,
and the signing key can sign an arbitrary message, but only once;
no QPT adversary can output a verification key of its own choice
together with valid signatures on two distinct messages under it.
OSS imply quantum lightning,
and quantum lightning implies public-key quantum money mini-schemes~\cite{STOC:AGKZ20,JC:Zhandry21}.
OSS have recently been constructed relative to a classical oracle,
and also in the plain model assuming sub-exponentially-secure
indistinguishability obfuscation and LWE~\cite{C:ShmZha25,EPRINT:HuaVai25}.

These public-key unclonable primitives easily imply
efficiently-verifiable state puzzles (EV-StatePuzzs).
An EV-StatePuzz is a pair $(\mathsf{Samp},\mathsf{Ver})$ of QPT algorithms with
the following syntax: $\mathsf{Samp}(1^\lambda)$ outputs a classical bit string
$s$, called the puzzle, and a quantum state $\psi_s$, called the answer, and
$\mathsf{Ver}(1^\lambda,s,\rho)$ accepts or rejects a candidate answer $\rho$.
Correctness requires that the honest answer $\psi_s$ is accepted with
overwhelming probability, and security requires that no QPT adversary, given
only the puzzle $s$, can output a state that is accepted by
$\mathsf{Ver}(1^\lambda,s,\cdot)$. 
EV-StatePuzzs are considered as ``zero-copy-secure'' public-key quantum money mini schemes:
in public-key quantum money mini schemes, a single copy of a quantum bank note is given to the adversary, and the adversary
has to generate another copy, while in EV-StatePuzzs, zero copy of a quantum bank note is given to the adversary.

Despite the intensive study of the constructions of public-key quantum money and its variants,
surprisingly little is known about their cryptographic consequences:
the known lower bounds of these unclonable primitives (and of EV-StatePuzzs)
are only one-way puzzles (OWPuzzs) and commitments~\cite{STOC:KhuTom24,STOC:KhuTom25}.
It is therefore natural to ask the following question.
\begin{center}
\emph{Are there any other useful cryptographic primitives that can be constructed from
one-shot signatures, quantum lightning, public-key quantum money, or EV-StatePuzzs,
beyond commitment schemes?}
\end{center}

\subsection{Our results}
In this paper, we show the following:
\begin{theorem}
\label{thm:main}
The following four primitives are existentially equivalent:
\begin{enumerate}
    \item 
    Everlasting-secure 2QKE.
\item 
Standard 2QKE.
    \item 
    Efficiently-verifiable state puzzles (EV-StatePuzzs).
    \item 
    One-time secure digital signatures with quantum signatures and quantum signing keys.
\end{enumerate}
\end{theorem}
The 2QKE that we construct has the following syntax: Alice and Bob are QPT. Alice sends a classical bit string
to Bob
over an authenticated classical channel, and a quantum state over an unauthenticated quantum channel.
Bob returns a quantum state to Alice over an unauthenticated quantum channel.
Alice outputs a bit $\alpha$ and Bob outputs a bit $\beta$.
Correctness requires that the probability that $\alpha=\beta$ is at least $1-\negl$.
The standard security requires that for any QPT Eve, who can freely read the classical message
and can freely modify the quantum messages,
the probability that $e=\beta$ is at most $\frac{1}{2}+\negl$, where $e$ is Eve's output.
Everlasting security considers the setup where Eve is QPT during the protocol,
but after that Eve can do any unbounded operation on her state to output $e$.
Everlasting-secure 2QKE implies 2QKE with the standard security.

\cref{thm:main} shows, in particular, that 2QKE are equivalent to EV-StatePuzzs. This
gives the first complete characterization of 2QKE.
Moreover, EV-StatePuzzs 
can be trivially constructed from OWFs,
but  
can exist even if $\mathsf{P}=\mathsf{NP}$ relative to a classical oracle~\cite{STOC:KreQiaTal25}.
Hence 
\begin{corollary}
\label{coro:PNP}
Relative to a classical oracle, $\mathsf{P}=\mathsf{NP}$ but everlasting-secure 2QKE exist.    
\end{corollary}
This means, in particular, that 2QKE are potentially weaker than OWFs.
Note that \cite{cryptoeprint:2024/1707} constructed a unitary oracle relative to which $\mathsf{BQP}=\mathsf{QCMA}$ but quantum multiparty
non-interactive key exchange over classical channel exist. 
Their KE is stronger than our 2QKE, but their oracle is a quantum one.

\cref{thm:main} also shows the equivalence between standard 2QKE and everlasting 2QKE.
This means that everlasting security comes for free: 
\begin{corollary}
If 2QKE exist, then everlasting-secure 2QKE exists.
\end{corollary}

In the proof of \cref{thm:main}, we construct everlasting-secure 2QKE from one-time secure digital signatures with quantum signatures.
Everlasting 2QKE thus constructed actually satisfies the special property: Alice does not need any quantum memory.
In other words, the internal state that Alice keeps between her first and second phases is classical. 
Hence we have the following corollary:
\begin{corollary}
If 2QKE exist, then everlasting-secure 2QKE in which Alice does not need quantum memory exists.
\end{corollary}

EV-OWPuzzs imply EV-StatePuzzs. Moreover,
planted QCMA~\cite{TCC:KitNisYam23} is equivalent to EV-OWPuzzs. Therefore
\begin{corollary}
If EV-OWPuzzs (or planted QCMA) exist, then everlasting-secure 2QKE exists.    
\end{corollary}

One-shot signatures~\cite{STOC:AGKZ20}, quantum lightning~\cite{JC:Zhandry21},
and public-key quantum money mini-schemes~\cite{STOC:AarChr12} imply EV-StatePuzzs.
Hence we also have the following corollary.
\begin{corollary}
\label{coro:unclonable_lowerbound}
If one-shot signatures, quantum lightning, or public-key quantum money mini-schemes exist, 
then everlasting-secure 2QKE exists.    
\end{corollary}
This gives a non-trivial lowerbound of these public-key unclonable primitives.
Whether the lowerbound is tight or not is open. In particular, showing that 2QKE imply public-key quantum money mini-schemes
will be very hard, because it also shows that OWFs imply public-key quantum money mini-schemes. 

\cite{cryptoeprint:2024/1707} constructed a unitary oracle relative to which $\mathsf{BQP}=\mathsf{QMA}$ but
quantum lightning exists. Since quantum lightning implies 2QKE from \cref{coro:unclonable_lowerbound}, we have the following corollary.
\begin{corollary}
Relative to a unitary oracle,
$\mathsf{BQP}=\mathsf{QMA}$
but
everlasting-secure 2QKE exist.
\end{corollary}

Finally, we also show the following:
\begin{theorem}
If one-shot signatures exist, then everlasting-secure QPKE with quantum public keys exist.    
\end{theorem}
Because one-shot signatures seem to be incomparable to OWFs, this result seems to imply that
(even everlasting-secure) QPKE with quantum public keys can be constructed from assumptions
incomparable to OWFs.

Moreover, QPKE constructed from one-shot signatures has an additional functionality, 
which we call \emph{certified freeze}, that allows the receiver to publish a classical 
certificate with which anyone can verify that the receiver has relinquished 
the ability of issuing new public keys.
\begin{corollary}
If one-shot signatures exist, then everlasting-secure QPKE with quantum public keys that satisfies certified freeze exist.    
\end{corollary}

\begin{figure}[H]
\centering
\begin{tikzpicture}[
  box/.style={draw, rounded corners, align=center, inner sep=5pt, minimum height=2.2em},
  arr/.style={-{Latex}, thick},
  dbl/.style={{Latex}-{Latex}, thick},
  nimp/.style={-{Latex}, thick,
    postaction={decorate, decoration={markings,
      mark=at position #1 with {\draw[-, thick] (-2.5pt,-5pt) -- (2.5pt,5pt);}}}},
  nimp/.default=0.3,
  lab/.style={fill=white, inner sep=1.5pt, font=\footnotesize},
  node distance=10mm and 18mm
]
\node[box] (oss)  {One-shot signatures};
\node[box] (ql)   [below=of oss] {Quantum lightning};
\node[box] (qm)   [below=of ql]  {Public-key quantum money\\ mini-schemes};
\node[box] (evsp) [below=36mm of qm]  {EV-StatePuzzs};

\node[box] (qke)  [right=of evsp, xshift=16mm] {2QKE};
\node[box] (ots)  [below=of evsp] {One-time signatures\\ with quantum signatures};
\node[box] (eqke) at (qke |- ots) {Everlasting-secure\\ 2QKE};
\node[box] (qpke) [above=of qke] {QPKE with\\ quantum pk};

\node[box] (evowp) [left=of evsp, xshift=-24mm, yshift=17mm] {EV-OWPuzzs};
\node[box] (owf)   [above=of evowp] {OWFs};
\node[box] (owp)   at (evowp |- ots) {OWPuzzs};
\node[box] (sp)    at ($(owp)!0.45!(evsp |- owp)$) {StatePuzzs};

\draw[arr] ([xshift=-3mm]oss.south) -- node[lab, left] {\cite{STOC:AGKZ20}} ([xshift=-3mm]ql.north);
\draw[nimp] ([xshift=3mm]ql.north) -- node[lab, right] {\cite{cryptoeprint:2024/1707}} ([xshift=3mm]oss.south);
\draw[arr] ([xshift=-3mm]ql.south) -- node[lab, left] {\cite{JC:Zhandry21}} ([xshift=-3mm]qm.north);
\draw[nimp] ([xshift=3mm]qm.north) -- node[lab, right] {\cite{EC:CKMSTY26}} ([xshift=3mm]ql.south);
\draw[arr] ([xshift=-3mm]owf.south) -- ([xshift=-3mm]evowp.north);
\draw[nimp] ([xshift=3mm]evowp.north) -- node[lab, right] {\cite{STOC:KreQiaTal25}} ([xshift=3mm]owf.south);
\draw[arr] (evowp.east) -| ([xshift=-5mm]evsp.north);
\draw[nimp=0.12] ([xshift=-10mm]evsp.north) |- node[lab, pos=0.75, below] {\cite{C:ChuGolGra24,cryptoeprint:2024/1707}} ([yshift=-3mm]evowp.east);
\draw[arr] ([xshift=-3mm]evowp.south) -- ([xshift=-3mm]owp.north);
\draw[nimp] ([xshift=3mm]owp.north) -- node[lab, right] {\cite{C:ChuGolGra24}} ([xshift=3mm]evowp.south);
\draw[arr] (evsp) -- (sp);
\draw[dbl] (sp) -- node[lab, below=2pt] {\cite{STOC:KhuTom25}} (owp);
\draw[arr] (eqke) -- (qke);
\coordinate (ossl) at ([xshift=-12mm]oss.west);
\draw[arr] (owf.east)
  -- ($(ossl |- owf.east)+(-1.2mm,0)$) arc[start angle=180, end angle=0, radius=1.2mm]
  -- ($(qm.south |- owf.east)+(-1.2mm,0)$) arc[start angle=180, end angle=0, radius=1.2mm]
  -| node[lab, pos=0.25, above] {\cite{C:KMNY24}} ([xshift=-3mm]qpke.north);

\draw[arr] (qm)   -- node[lab, right, pos=0.68] {\cref{lem:qots-from-evsp}} (evsp);
\draw[arr] (evsp) -- node[lab, right] {\cref{lem:qots-from-evsp}} (ots);
\draw[arr] (ots)  -- node[lab, above] {\cref{thm:everlasting-ke}} (eqke);
\draw[arr] (qke)  -- node[lab, above] {\cref{lem:2QKE-to-gap}} (evsp);
\draw[arr] (oss.east) -| node[lab, pos=0.25, above] {\cref{thm:OSS_to_QPKE}} ([xshift=3mm]qpke.north);
\draw[arr] (oss.west) -- ++(-12mm,0) |- node[lab, pos=0.25, left] {} ([yshift=2.5mm]evowp.east);
\draw[arr] (qpke) -- (qke);
\end{tikzpicture}
\caption{
Summary of known results and our results.
Note that the constructions of QPKE from
OWFs and one-shot signatures achieve slightly different properties.
(For the difference, see \cref{sec:QPKE_difference}.)
In this figure, for simplicity, we do not distinguish them.
}
\label{fig}
\end{figure}
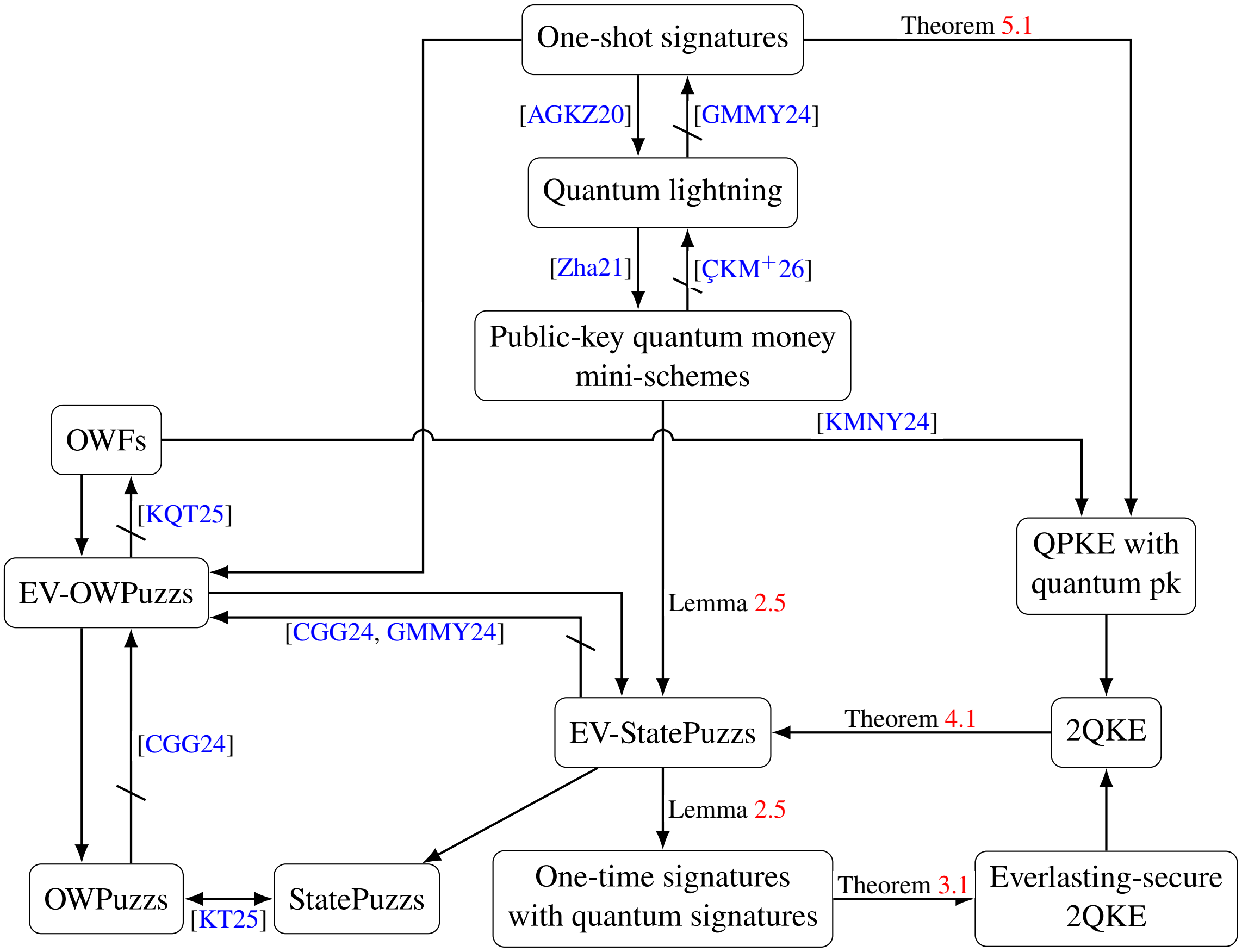

The known results and our results are summarized in \cref{fig}.

\subsection{Technical Overview}
In this subsection, we provide high-level overview
of our proofs.

\paragraph{From EV-StatePuzzs to one-time signatures.}
This is simple.
The construction is a quantum analogue of Lamport signatures.
$\mathsf{KeyGen}$ runs $(s_b,\psi_b)\leftarrow\mathsf{Samp}(1^\lambda)$ for each $b\in\{0,1\}$,
and outputs $\vk\coloneqq(s_0,s_1)$ and $\sigk\coloneqq(\psi_0,\psi_1)$.
The signature of $m\in\{0,1\}$ is $\psi_m$, and the verification of $(m,\sigma)$ runs $\mathsf{Ver}(1^\lambda,s_m,\sigma)$.
For one-time security, the reduction embeds its challenge puzzle $s^*$ at a random position $b^*\leftarrow\{0,1\}$ of $\vk$
and samples the other puzzle--answer pair by itself.
With probability $1/2$, the adversary queries the signature of $m=1-b^*$, which the reduction can answer with its self-sampled state,
and the forged signature for $1-m=b^*$ is a state accepted by $\mathsf{Ver}(1^\lambda,s^*,\cdot)$, i.e., an answer to the challenge puzzle.

\paragraph{From one-time signatures to everlasting-secure 2QKE.}
Before explaining our construction, we recall the construction of \cite{C:MalWal24,C:KMNY24}
where digital signatures with \emph{classical} signatures are used. 

Let $(\mathsf{KeyGen},\mathsf{Sign},\mathsf{Vrfy})$ be a one-time secure digital signature scheme with classical signatures.
Alice runs $(\mathsf{vk},\mathsf{sigk})\leftarrow\mathsf{KeyGen}(1^\lambda)$ and generates the signatures $\sigma(0)$ and $\sigma(1)$ of both messages.
She prepares the state
\begin{align}
|0\rangle_\regA|\sigma(0)\rangle_\regB
+(-1)^a|1\rangle_\regA|\sigma(1)\rangle_\regB,
\label{eq:overview_state}
\end{align}
where $a\leftarrow\{0,1\}$ is a uniformly random bit. 
She sends the registers $(\regA,\regB)$ to Bob over the unauthenticated quantum channel, and sends $\mathsf{vk}$ over the authenticated classical channel.
Bob coherently checks that the register $\regB$ contains a valid signature of the message stored in the register $\regA$.
If the check rejects, Bob outputs a uniformly random bit $\beta$ and returns the maximally mixed state to Alice.
If the check accepts, the state is only negligibly disturbed, and Bob samples $b\leftarrow\{0,1\}$, applies $Z^b$ on the register $\regA$, returns the state to Alice, and outputs $\beta\coloneqq b$.
Alice's state is now
\begin{align}
|0\rangle_\regA|\sigma(0)\rangle_\regB
+(-1)^{a\oplus b}|1\rangle_\regA|\sigma(1)\rangle_\regB,
\end{align}
from which she can easily learn $a\oplus b$, because she knows both $\sigma(0)$ and $\sigma(1)$.
She can therefore learn $b$.

The security is intuitively understood in the following way:
Because the relative phase $a$ is unknown to Eve,
Alice's first phase is indistinguishable from the following one
to Eve: 
Alice chooses $c\gets\bit$ and sends $|c\rangle|\sigma(c)\rangle$ to Bob.
By using $|c\rangle|\sigma(c)\rangle$, Eve generates a state $\rho$ and sends it to Bob.
Bob performs the test on $\rho$. Assume that the test passes, and
let $\rho'$ be the state after the test.
Assuming the strong existential unforgeability\footnote{The strong existential unforgeability means that 
no QPT adversary can
produce a different signature (even on the same message) than the one provided by
the signing oracle.},
we can show that $\rho'$ is statistically close to $|c\rangle|\sigma(c)\rangle$.
This is because, if $\rho'$ contains a valid-message-signature-pair state $|d\rangle|\tau(d)\rangle$ such that
$d=1-c$ or $d=c\wedge\tau(c)\neq\sigma(c)$,
then $\rho$ also contains it, which contradicts the strong existential unforgeability.
Then, even if Bob applies $Z^b$ on $\rho'$, no information about $b$ is encoded on $\rho'$,
and therefore even unbounded Eve cannot learn $b$ from the state sent from Bob.

This idea, however, does not work if the signatures are quantum states.
In general, the signing algorithm for quantum signatures works as follows: it applies a unitary $U$ on $|0...0\rangle$ to
generate the state $\sum_i C_i |i\rangle|\phi_i\rangle$,
and measure the first register in the computational basis.
If the result is $i$, the second register collapses to $|\phi_i\rangle$.
Some qubits of $|\phi_i\rangle$ are output as the quantum signature.
Without loss of generality, we can assume that quantum signatures are pure, because
we can redefine $(I\otimes P)|\phi_i\rangle$ as the signature, where $P$ is a random Pauli operator whose description is stored in the signing key.

For such a signing algorithm, the idea of \cite{C:MalWal24,C:KMNY24} that we have explained above does not work.
First, even if Alice generates two quantum signatures, $|\sigma(0)\rangle,|\sigma(1)\rangle$, by running the signing algorithm twice,
she cannot generate the superposition
\begin{align}
|0\rangle|\sigma(0)\rangle+|1\rangle|\sigma(1)\rangle    
\end{align}
from $|\sigma(0)\rangle\otimes|\sigma(1)\rangle$.
Second, the signing algorithm is not necessarily deterministic, and therefore Alice cannot learn
$a\oplus b$ from $|0\rangle|\sigma(0)\rangle+(-1)^{a\oplus b}|1\rangle|\sigma(1)\rangle$.
Third, we cannot assume the strong existential unforgeability, because in the first place
it is not well defined for quantum signatures.

The first two issues are resolved by modifying Alice's first phase as follows:
\begin{enumerate}
    \item 
    Generate $|+\rangle|\sigma(0)\rangle|\sigma(1)\rangle$.
    \item 
    Apply the controlled-SWAP to get
    $|0\rangle|\sigma(0)\rangle|\sigma(1)\rangle+|1\rangle|\sigma(1)\rangle|\sigma(0)\rangle$.
    \item 
    Apply a random Pauli $P$ to the third register to get
    $|0\rangle|\sigma(0)\rangle P|\sigma(1)\rangle+|1\rangle|\sigma(1)\rangle P|\sigma(0)\rangle$.
\end{enumerate}
Because $P$ is unknown to Eve, the third register is ``masked'' and Alice's state ``looks like''
    $|0\rangle|\sigma(0)\rangle+|1\rangle|\sigma(1)\rangle$.
Moreover, Alice can map
    $|0\rangle|\sigma(0)\rangle P|\sigma(1)\rangle+(-1)^{a\oplus b}|1\rangle|\sigma(1)\rangle P|\sigma(0)\rangle$
    to
    $(|0\rangle+(-1)^{a\oplus b}|1\rangle)|\sigma(0)\rangle |\sigma(1)\rangle$
by applying $P$ and the controlled-SWAP again,
and from the latter state she can learn $a\oplus b$.

To resolve the third issue, we had to construct a different proof than those of \cite{C:KMNY24,C:MalWal24}.
Because we cannot assume strong existential unforgeability, we cannot say that the state $\rho'$ after Bob's test
is $|c\rangle|\sigma(c)\rangle$. However, still we can say that $\rho'$
is spanned by valid-message-signature-pair states $\{|d\rangle|\tau(d)\rangle\}$. 
In order for Eve to get the information about $b$, 
$\rho'$ (and therefore the state $\rho$ Eve sends to Bob) must contain high weight
on both $|0\rangle|\tau(0)\rangle$ and $|1\rangle|\tau(1)\rangle$. However,
this means Eve has the ability to forge a signature for the message $1-c$, which contradicts the security of the signature scheme.

\paragraph{From 2QKE to EV-StatePuzzs.}
Finally, we explain the converse direction. 
\cite{EC:QiaRaiZha25} and \cite{cryptoeprint:2024/1707} already showed that
2QKE imply StatePuzzs, but we observe that
we can actually construct
EV-StatePuzzs.

By the principle of deferred measurement, we can assume without loss of generality that Alice's first phase
has the following form: she applies a unitary $U$ on $|0...0\rangle$ over three registers
$(\mathbf{S},\mathbf{M},\mathbf{W})$ to generate the state
\begin{align}
U|0...0\rangle=\sum_{s}C_s|s\rangle_{\mathbf{S}}|\Phi_s\rangle_{\mathbf{M},\mathbf{W}},
\end{align}
measures $\mathbf{S}$ in the computational basis to obtain $s$, sends $s$ over the authenticated classical
channel and the register $\mathbf{M}$ over the unauthenticated quantum channel, and keeps the register
$\mathbf{W}$.

We first construct a $1/2$-sound EV-StatePuzz from a 2QKE. Here, a
$\delta$-sound EV-StatePuzz is an EV-StatePuzz where any QPT adversary can make the verification algorithm accept with probability
at most $\delta+\negl$.
The construction is as follows.
$\mathsf{Samp}(1^\lambda)$ runs Alice's first phase, and outputs the puzzle $s$ and the answer
$\psi_s\coloneqq|\Phi_s\rangle$, which is the state of the registers $(\mathbf{M},\mathbf{W})$.
$\mathsf{Ver}(1^\lambda,s,\rho)$ runs Bob's honest algorithm on $s$ and the register $\mathbf{M}$ of $\rho$,
which outputs a bit $\beta$ and a returned quantum state, then runs Alice's second phase on $s$, the register
$\mathbf{W}$ of $\rho$, and the returned state, which outputs a bit $\alpha$, and finally accepts if and only
if $\alpha=\beta$.
The correctness of the EV-StatePuzz is exactly the correctness of the 2QKE.
For the security, let us consider a QPT adversary $\cA$, given $s$, outputs a state that is accepted by
$\mathsf{Ver}(1^\lambda,s,\cdot)$ with probability at least $1/2+1/\poly$.
Then Eve can run $\cA$ on input $s$, and send the $\regM$ register of the output state to Bob.
Eve then simulates Alice's second phase on her state and the state returned from Bob.
Conditioned on each $s$, the experiment of $\mathsf{Ver}(1^\lambda,s,\cA(1^\lambda,s))$ is identical to the
attacked execution of the 2QKE, where the acceptance condition $\alpha=\beta$ corresponds to the event that
Eve's output $e$ agrees with Bob's output $\beta$.
Hence 
Eve breaks the security of the 2QKE.

The $1/2$-security of the EV-StatePuzz can be amplified by parallel repetition~\cite{STOC:BQSY24},
and finally we obtain an EV-StatePuzz with $\negl$ security.

\paragraph{QPKE from one-shot signatures.}
Finally, we explain our construction of everlasting-secure QPKE with quantum public keys from one-shot signatures (OSS)~\cite{STOC:AGKZ20}.
An OSS scheme has a classical common reference string $\mathsf{crs}$, a classical verification key, a quantum signing key, classical signatures, and a classical deterministic verification algorithm.
Its security requires that no QPT adversary, given $\mathsf{crs}$, can output a verification key of its own choice together with valid signatures on two distinct messages under it.
In particular, a signing key can be used only once, even by its honest holder.

Recall the 2QKE schemes of~\cite{C:MalWal24,C:KMNY24} explained above.
Alice's message, namely a classical $\mathsf{vk}$ and the state $\ket{0}\ket{\sigma(0)}+(-1)^a\ket{1}\ket{\sigma(1)}$, can be regarded as a quantum public key.
In the QPKE setting, however, the receiver has to distribute unbounded polynomially many public keys, all of which must be verifiable under the single authenticated $\mathsf{vk}$, and this is why \cite{C:MalWal24,C:KMNY24} need many-time secure digital signatures, which are available from OWFs.
Our 2QKE from EV-StatePuzzs does not directly extend to QPKE for the same reason.
OSS is also a ``one-time'' object, but its signatures are classical.
This allows us to apply the certificate-chain technique, which is a classical folklore technique to convert one-time signatures into stateful many-time signatures~\cite{C:Merkle87,C:Merkle89a,STOC:NaoYun89,SIAMCOMP:GolMicRiv88}: the $t$-th one-time key signs the $t$-th message together with a freshly generated $(t+1)$-th verification key, and the $t$-th signature consists of the resulting chain of certificates.
Because the certificates are classical, they can be freely copied into every subsequent public key.

Our construction of QPKE from OSS is as follows.
$\mathsf{KeyGen}$ runs $\mathsf{crs}\leftarrow\mathsf{Setup}(1^\lambda)$ and $(w_1,\mathsf{sigk}_1)\leftarrow\mathsf{Gen}(\mathsf{crs})$, outputs $\mathsf{vk}\coloneqq(\mathsf{crs},w_1)$, and keeps $\mathsf{sigk}_1$ as the initial secret key.
On the $t$-th invocation, $\mathsf{PKGen}$ generates three fresh key pairs $(u^0_t,\mathsf{sigk}^0_t)$, $(u^1_t,\mathsf{sigk}^1_t)$, and $(w_{t+1},\mathsf{sigk}_{t+1})$, signs $\mu_t\coloneqq u^0_t\|u^1_t\|w_{t+1}$ with the chain key $\mathsf{sigk}_t$ to obtain the signature $\tau_t$, and, for each $\alpha\in\{0,1\}$, signs the message $\alpha$ with the key $\mathsf{sigk}^\alpha_t$ to obtain a classical signature $\sigma_t(\alpha)$.
It then samples $a_t\leftarrow\{0,1\}$, prepares
\begin{align}
\frac{1}{\sqrt{2}}\left(\ket{0}_{\mathbf{A}}\ket{\sigma_t(0)}_{\mathbf{B}}+(-1)^{a_t}\ket{1}_{\mathbf{A}}\ket{\sigma_t(1)}_{\mathbf{B}}\right),
\end{align}
and outputs the public key $\mathsf{pk}_t$ consisting of $t$, the chain $(\mu_s,\tau_s)_{s\le t}$, and this state.
It stores $(a_t,\sigma_t(0),\sigma_t(1))$ classically and keeps $\mathsf{sigk}_{t+1}$ for the next invocation.
$\mathsf{Enc}(\mathsf{vk},\mathsf{pk}_t,m)$ verifies the chain starting from $w_1$, coherently checks that the register $\mathbf{B}$ contains a valid signature under $u^{\alpha}_t$ of the message $\alpha$ stored in the register $\mathbf{A}$, applies $Z^m$ on $\mathbf{A}$, measures $(\mathbf{A},\mathbf{B})$ in the Hadamard basis to obtain $(c,d)$, and outputs the classical ciphertext $(t,c,d)$.
Because the receiver knows $a_t$, $\sigma_t(0)$, and $\sigma_t(1)$, it recovers $m=c\oplus a_t\oplus d\cdot(\sigma_t(0)\oplus\sigma_t(1))$.
Note that, because OSS signatures are classical, the receiver can directly prepare the superposition of the two signatures and knows both of them, and the sender can send the classical Hadamard-basis measurement outcome as in~\cite{C:MalWal24,C:KMNY24}; the controlled-SWAP and the random Pauli used in our 2QKE are not needed, and ciphertexts are classical.
On the other hand, unlike \cite{C:MalWal24,C:KMNY24}, a single OSS key cannot sign both $0$ and $1$, and this is why each public key has two verification
keys $u^0_t,u^1_t$, which are signed together with the next chain key.

Although the chain itself is the classical one, both its role and its analysis differ from the classical setting.
First, in the classical setting the chain is used to sign messages, whereas here it is used only to certify the keys, and the ``payload'' of each public key is a quantum state, namely a superposition of two classical signatures whose relative phase $a_t$ is a secret of the receiver.
Second, in the classical setting the security of the chain is reduced to one-time unforgeability under a challenge verification key given by the challenger, and the reduction has to guess the position at which the adversary deviates from the honest chain.
In our setting the OSS adversary chooses the verification key by itself and has no signing oracle, and therefore our reductions simply run the entire security game by themselves, keep all honest keys, and output two colliding signatures; no guessing is involved.
Third, only the weak form of unforgeability, in which the two forged messages must be distinct, is needed.
This is in contrast to \cite{C:MalWal24,C:KMNY24}, whose proofs rely on strong unforgeability to argue that the state after the coherent verification is exactly $\ket{c}\ket{\sigma(c)}$; instead, we reuse the ``cross-term'' argument of our 2QKE proof explained above.

The security proof goes as follows.
Since $\mathsf{Dec}$ is never run in the security game, averaging over the pad $a_t$ dephases $\mathbf{A}$, and hence the adversary's view is unchanged if the challenger samples $\alpha_t\leftarrow\{0,1\}$ and sends $\ket{\alpha_t}\ket{\sigma_t(\alpha_t)}$ instead, without ever signing with $\mathsf{sigk}^{1-\alpha_t}_t$; the key $\mathsf{sigk}^{1-\alpha_t}_t$ thus remains unconsumed in the challenger's hand.
This replacement is an equality of states, and is therefore valid even against the unbounded phase of the adversary.
Suppose that the adversary has made $q'$ public-key queries before the challenge, so that the challenger has issued the honest chain $(\mu_1,\tau_1),\ldots,(\mu_{q'},\tau_{q'})$ and holds the next chain signing key $\mathsf{sigk}_{q'+1}$, corresponding to the verification key $w_{q'+1}$ contained in $\mu_{q'}$ (for $q'=0$, this is $w_1$), which has not signed anything yet.
The adversary then submits a public key, whose classical part is measured to obtain a chain $(\mu'_1,\tau'_1),\ldots,(\mu'_{t'},\tau'_{t'})$ of some length $t'$.
If this chain does not verify under $w_1$, the challenge ciphertext is $\bot$, and there is nothing to prove.
Otherwise, we distinguish two cases according to whether the submitted chain is a prefix of the honest chain, i.e., whether $t'\le q'$ and $\mu'_s=\mu_s$ for all $s\le t'$.

If the submitted chain is not a prefix of the honest chain, then the adversary has forged a certificate under an honestly generated chain key.
Indeed, let $r$ be the first position where $\mu'_r\ne\mu_r$.
If $r\le q'$, then $\tau'_r$ is a valid signature of $\mu'_r$ under $w_r$, and the challenger already holds another valid signature $\tau_r$ of the different message $\mu_r$ under the same key $w_r$.
If no such $r\le q'$ exists, then $t'>q'$ and $\tau'_{q'+1}$ is a valid signature of $\mu'_{q'+1}$ under $w_{q'+1}$; since $\mathsf{sigk}_{q'+1}$ is unconsumed, the challenger can sign a different message with it.
In both cases, a reduction that simulates the challenger obtains two valid signatures on distinct messages under an honestly generated verification key, which contradicts the security of OSS.
Hence this case occurs only with negligible probability.

If the submitted chain is a prefix of the honest chain, then the keys used by $\mathsf{Enc}$ are the honest ones, $(u^0_j,u^1_j)$ with $j\coloneqq t'$.
Recall that, in the modified security game, the challenger sent $\ket{\alpha_j}\ket{\sigma_j(\alpha_j)}$ and still holds the unconsumed leaf key $\mathsf{sigk}^{1-\alpha_j}_j$.
As in the proof of our 2QKE, after the coherent check the submitted state on the registers $(\mathbf{A},\mathbf{B})$ is spanned by valid pairs $\ket{\alpha}\ket{\beta}$ with $\top=\mathsf{Ver}(\mathsf{crs},u^\alpha_j,\alpha,\beta)$, and applying $Z^b$ on $\mathbf{A}$ can encode $b$ only in the cross term between the $\ket{0}$-branch and the $\ket{1}$-branch.
The weight of this cross term is bounded by the weight of the branch $\ket{1-\alpha_j}\ket{\beta}$ with $\beta$ a valid signature of $1-\alpha_j$ under $u^{1-\alpha_j}_j$, and such a $\beta$ is a forgery: measuring $(\mathbf{A},\mathbf{B})$ in the computational basis to obtain it and signing the other message $\alpha_j$ with the unconsumed $\mathsf{sigk}^{1-\alpha_j}_j$ yields two valid signatures on distinct messages under $u^{1-\alpha_j}_j$, which again contradicts the security of OSS.
Hence the cross term has negligible weight, and the trace distance between the adversary's states for $b=0$ and $b=1$ is negligible.

\subsection{Remarks on Our 2QKE}
\label{sec:remarks_2QKE}
Here we provide several remarks on our 2QKE.

First, the 2QKE scheme constructed from EV-StatePuzzs in this paper is a special version of 2QKE where Bob can choose his own key.
In general 2QKE, Bob's key is randomly generated and he cannot choose his key.
However, in our 2QKE scheme, Bob can choose his own key. (This is clear from our construction of 2QKE.)
The 2QKE schemes constructed in \cite{C:MalWal24,C:KMNY24} from OWFs also satisfy this property.
On the other hand, our lowerbound, 2QKE $\Rightarrow$ EV-StatePuzzs, holds 
for general 2QKE where Bob cannot choose his key. (This is also clear from the construction of EV-StatePuzzs.)

Second, in the 2QKE schemes of \cite{C:MalWal24,C:KMNY24}, Bob's message can be classical. Because they use digital signatures with \emph{classical} signatures,
the $\regB$ register of the state $|0\rangle_\regA|\sigma(0)\rangle_\regB+(-1)^{a\oplus b}|1\rangle_\regA|\sigma(1)\rangle_\regB$ is
in the computational basis state.
Hence, instead of sending the state, Bob can send the Hadamard-basis measurement result on the state from which Alice can
recover $a\oplus b$.
Because we use digital signatures with \emph{quantum} signatures, the state of the $\regB$ register is not necessarily in the computational-basis state,
and therefore we cannot use the same technique.
It is an open problem whether we can construct 2QKE with classical Bob's message from EV-StatePuzzs or other assumptions weaker than OWFs.

Third, in our 2QKE, we require security only for Bob: we require that Eve cannot learn Bob's key.
On the other hand, we do not require security for Alice, and therefore Eve could learn Alice's key.
In fact, it is unavoidable with the current syntax, because Eve can impersonate Bob: Eve can take Alice's message and run Bob's algorithm,
and return the message to Alice. Then due to the correctness, Eve's key agrees with Alice's key.
The 2QKE schemes of \cite{C:MalWal24,C:KMNY24} do not have this issue because Bob's message is classical, and therefore
it can be authenticated: Eve cannot impersonate Bob. 
Because Bob's message is quantum in our 2QKE scheme, we cannot use the same idea.
However, actually, there is a solution
if Bob additionally sends a short classical bit string over the authenticated classical channel.
(For details, see \cref{sec:strong}.)

Fourth, in our 2QKE, Eve can make Alice accept a key that is different from Bob's key in the following way: 
Eve does not touch Alice's message to Bob, but Eve applies $Z^d$ with random $d\gets\bit$
on the $\regA$ register of the state sent from Bob to Alice so that Alice outputs her key
$b\oplus d$.
Again, the 2QKE schemes of \cite{C:MalWal24,C:KMNY24} do not have this issue because Bob's message is classical and authenticated.
However, we can again address the issue in a similar way.
(For details, see \cref{sec:strong}.)

Finally,
\cite{C:KMNY24} constructed quantum PKE with quantum public keys that are secure even if
unbounded polynomially-many public keys are distributed.
This is possible because they use OWFs, which allow many-time secure digital signatures.
On the other hand, we do not know how to obtain many-time secure digital signatures from 
EV-StatePuzzs.
Constructing quantum PKE (even with quantum public keys) from a weaker assumption
than OWFs remains open.

\subsection{Remarks on Our QPKE}
\label{sec:QPKE_difference}
Unlike \cite{C:KMNY24}, $\mathsf{PKGen}$ of our QPKE constructed from OSS is stateful, and 
the size of the classical part of the public key grows with the number of uses.
Moreover, the secret key is quantum. The size of the classical part of the secret key grows with the number of uses.
On the other hand, our QPKE achieves everlasting security.

The length of $\mathsf{pk}_t$ grows linearly in $t$ because the chain has length $t$.
A standard Merkle-tree variant, in which the $t$-th leaf is certified by a path of length $\lambda$ in a lazily generated binary tree, reduces it to $O(\lambda)$ times the length of a verification key and a signature, at the cost of storing $O(\lambda)$ chain signing keys in the secret key and a slightly more involved case analysis in the proof.

\subsection{Related Works}
StatePuzzs from public-key quantum money was shown in \cite{EC:QiaRaiZha25,STOC:KhuTom25}. 
StatePuzzs from 2QKE was shown in \cite{EC:QiaRaiZha25,cryptoeprint:2024/1707}.
In this paper, we observed that actually EV-StatePuzzs are implied.

\section{Preliminaries}

\subsection{Public-Key Quantum Money}
\begin{definition}[Public-key quantum money mini-scheme~\cite{STOC:AarChr12}]\label{def:minischeme}
A public-key quantum money mini-scheme consists of two QPT algorithms $(\mathsf{Mint},\mathsf{Vrfy})$ with the following syntax.
\begin{itemize}
    \item $\mathsf{Mint}(1^\lambda)\to(s,\rho_s)$: The minting algorithm takes the security parameter $1^\lambda$ as input, and outputs a classical serial number $s$ and a quantum state $\rho_s$.
    \item $\mathsf{Vrfy}(1^\lambda,s,\rho)\to\top/\bot$: The verification algorithm takes the security parameter $1^\lambda$, a serial number $s$, and a quantum state $\rho$ as input, and outputs $\top$ or $\bot$.
\end{itemize}
We require the following two properties.

\paragraph{Correctness.}
For all $\lambda\in\mathbb{N}$,
\begin{align}
\Pr\left[\top\leftarrow\mathsf{Vrfy}(1^\lambda,s,\rho_s) : (s,\rho_s)\leftarrow\mathsf{Mint}(1^\lambda)\right]\geq 1-\mathsf{negl}(\lambda).
\end{align}

\paragraph{Security.}
Define the algorithm $\mathsf{Vrfy}_2$ as follows: $\mathsf{Vrfy}_2(1^\lambda,s,\sigma)$ takes as input a serial number $s$ and a (possibly entangled) quantum state $\sigma$ over two registers, applies $\mathsf{Vrfy}(1^\lambda,s,\cdot)$ to each register, and outputs $\top$ if and only if both invocations output $\top$.
Then, for any QPT adversary $\mathcal{A}$,
\begin{align}
\Pr\left[\top\leftarrow\mathsf{Vrfy}_2(1^\lambda,s,\sigma) : 
\begin{array}{l}
(s,\rho_s)\leftarrow\mathsf{Mint}(1^\lambda)\\
\sigma\leftarrow\mathcal{A}(1^\lambda,s,\rho_s)
\end{array}
\right]\leq\mathsf{negl}(\lambda).
\end{align}
\end{definition}

\subsection{EV-StatePuzzs}
\begin{definition}[Efficiently verifiable state puzzles (EV-StatePuzzs)]
Let $\delta:\mathbb{N}\to[0,1]$.
A $\delta$-sound efficiently verifiable state puzzle (EV-StatePuzz) is a pair $(\Samp,\Ver)$ of QPT algorithms with the following syntax:
\begin{itemize}
\item
$\Samp(1^\lambda)\to(s,\psi_s)$:
It is a QPT algorithm that, on input the security parameter $\lambda$, outputs a classical bit string $s$ and a quantum state $\psi_s$.
$s$ is called the puzzle and $\psi_s$ is called answer.
\item
$\Ver(1^\secp,s,\rho)\to\top/\bot$:
It is a QPT algorithm that, on input $1^\secp$, $s$ and a quantum state $\rho$, outputs $\top$ or $\bot$.
\end{itemize}
We require the following two properties.
\begin{itemize}
\item
Correctness:
\begin{align}
\Pr[\top\gets\Ver(1^\secp,s,\psi_s):(s,\psi_s)\gets\Samp(1^\lambda)]\ge1-\mathsf{negl}(\lambda).
\end{align}
\item
Security:
For any QPT adversary $\mathcal{A}$,
\begin{align}
\Pr[\top\gets\Ver(1^\secp,s,\mathcal{A}(1^\lambda,s)):(s,\psi_s)\gets\Samp(1^\lambda)]\le\delta(\secp)+\mathsf{negl}(\lambda).
\end{align}
\end{itemize}
We call a $\delta$-sound EV-StatePuzz just EV-StatePuzz if $\delta=0$.
\end{definition}

\subsection{Digital Signatures}
\begin{definition}[One-time digital signatures with quantum signatures]\label{def:qots}
A one-time digital signature scheme consists of three QPT algorithms $(\mathsf{KeyGen},\mathsf{Sign},\mathsf{Ver})$ with the following syntax.
\begin{itemize}
    \item $\mathsf{KeyGen}(1^\lambda)\to(\mathsf{vk},\mathsf{sigk})$: The key generation algorithm takes the security parameter $1^\lambda$ as input, and outputs a classical verification key $\mathsf{vk}$ and a quantum signing key $\mathsf{sigk}$.
    \item $\mathsf{Sign}(\mathsf{sigk},m)\to\sigma$: The signing algorithm takes the signing key $\mathsf{sigk}$ and a message $m\in\{0,1\}$ as input, and outputs a quantum signature $\sigma$.
    \item $\mathsf{Ver}(\mathsf{vk},m,\sigma)\to\top/\bot$: The verification algorithm takes the verification key $\mathsf{vk}$, a message $m\in\{0,1\}$, and a quantum state $\sigma$ as input, and outputs $\top$ or $\bot$.
\end{itemize}
We require the following two properties.

\paragraph{Correctness.}
For all $m\in\{0,1\}$,
\begin{align}
\Pr\left[\top\leftarrow\mathsf{Ver}(\mathsf{vk},m,\sigma) :
\begin{array}{l}
(\mathsf{vk},\mathsf{sigk})\leftarrow\mathsf{KeyGen}(1^\lambda)\\
\sigma\leftarrow\mathsf{Sign}(\mathsf{sigk},m)
\end{array}
\right]\geq 1-\mathsf{negl}(\lambda).
\end{align}

\paragraph{One-time security.}
For any QPT adversary $\mathcal{A}=(\mathcal{A}_1,\mathcal{A}_2)$,
\begin{align}
\Pr\left[\top\leftarrow\mathsf{Ver}(\mathsf{vk},1-m,\sigma') :
\begin{array}{l}
(\mathsf{vk},\mathsf{sigk})\leftarrow\mathsf{KeyGen}(1^\lambda)\\
(m,\mathsf{st})\leftarrow\mathcal{A}_1(\mathsf{vk})\\
\sigma\leftarrow\mathsf{Sign}(\mathsf{sigk},m)\\
\sigma'\leftarrow\mathcal{A}_2(\sigma,\mathsf{st})
\end{array}
\right]\leq\mathsf{negl}(\lambda),
\end{align}
where $\mathsf{st}$ is a (possibly quantum) internal state of $\mathcal{A}_1$.
\end{definition}

\begin{remark}\label{rem:pure-signature}
Without loss of generality, we will assume that the signature $\sigma$ is a pure state determined exclusively by the signing key and the message. In particular, we can include the description of a random Pauli in the signing key, and then have the signature of a message $m$ be a purification of the mixed signature with the random Pauli applied to the purification register.
\end{remark}

\subsection{Constructing Digital Signatures}

We make the following observations, with a formal proof included in \cref{app:relations}. The first result follows from the observation that when looking at the construction of state puzzles from public-key quantum money mini-schemes, the verification algorithm of the mini-scheme gives a verification algorithm for the resulting puzzle. The second result follows from generalizing Lamport's construction of one-time signatures from one-way functions~\cite{Lamport_signature}.
\begin{lemma}
\label{lem:qots-from-evsp}
The following hold.
\begin{enumerate}
\item 
If there exist public-key quantum money mini-schemes,
then there exist EV-StatePuzzs.
    \item 
If there exist EV-StatePuzzs, 
then there exist one-time secure digital signatures.
\end{enumerate}
\end{lemma}

\subsection{2QKE}
\begin{definition}[Two-message quantum key exchange (2QKE)]
\label{def:2QKE}
A two-message quantum key exchange (2QKE) is the following interactive protocol between QPT Alice and QPT Bob.
Alice and Bob share a quantum channel, which is not authenticated, and a classical channel, which is authenticated.
\begin{enumerate}
\item 
Alice and Bob take $1^\secp$ as input.
    \item 
    Alice sends a quantum state and a classical bit string to Bob.
    \item 
    Bob returns a quantum state to Alice.
    \item 
    Alice outputs a bit $\alpha$ and Bob outputs a bit $\beta$.
\end{enumerate}
We require the following two properties.

\paragraph{Correctness.}
The probability that $\alpha=\beta$ is at least $1-\negl(\secp)$.

\paragraph{Security.}
For any QPT Eve who outputs a bit $e$, the probability that $e=\beta$ is at most $\frac{1}{2}+\negl(\secp)$. 
Here, Eve can freely read the classical messages and can freely modify the quantum messages.
\end{definition}

\begin{definition}[Everlasting security of 2QKE~\cite{C:Unruh13,C:MalWal24}]\label{def:everlasting}
A 2QKE satisfies everlasting security if the following holds for any Eve of the following form:
during the execution of the protocol, Eve can freely read the classical messages and can freely modify the quantum messages, but all of her operations are QPT;
after the protocol terminates, Eve applies computationally-unbounded operations on all the registers that she keeps, and outputs a bit $e$.
The requirement is that the probability that $e$ is equal to Bob's output $\beta$ is at most $\frac{1}{2}+\negl(\secp)$.
\end{definition}

\begin{remark}
Everlasting security implies the security of 2QKE, because a QPT Eve is a special case of the Eve of \cref{def:everlasting}.
\end{remark}

\subsection{One-Shot Signatures}
\begin{definition}[One-shot signatures~\cite{STOC:AGKZ20}]\label{def:OSS}
A one-shot signature scheme consists of four algorithms $(\mathsf{Setup},\mathsf{Gen},\mathsf{Sign},\mathsf{Ver})$ with the following syntax.
\begin{itemize}
\item
$\mathsf{Setup}(1^\lambda)\to\mathsf{crs}$:
The setup algorithm is a QPT algorithm that
takes the security parameter $1^\lambda$ as input, and outputs a classical common reference string $\mathsf{crs}$.
\item
$\mathsf{Gen}(\mathsf{crs})\to(\mathsf{vk},\mathsf{sigk})$:
The key generation algorithm is a QPT algorithm that
takes $\mathsf{crs}$ as input, and outputs a classical verification key $\mathsf{vk}$ and a quantum signing key $\mathsf{sigk}$.
\item
$\mathsf{Sign}(\mathsf{crs},\mathsf{sigk},m)\to\sigma$:
The signing algorithm is a QPT algorithm that
takes $\mathsf{crs}$, the signing key $\mathsf{sigk}$, and a message $m\in\{0,1\}^*$ as input, and outputs a classical signature $\sigma$.
\item
$\mathsf{Ver}(\mathsf{crs},\mathsf{vk},m,\sigma)\to\top/\bot$:
The verification algorithm is a classical deterministic polynomial-time algorithm
that takes $\mathsf{crs}$, the verification key $\mathsf{vk}$, a message $m\in\{0,1\}^*$, and a signature $\sigma$ as input, and outputs $\top$ or $\bot$.
\end{itemize}
We require the following two properties.

\paragraph{Correctness.}
For any quantum algorithm $M$ that outputs a classical message $m\in\{0,1\}^{*}$, which may be computationally unbounded,
\begin{align}
\Pr\left[
\top\leftarrow\mathsf{Ver}(\mathsf{crs},\mathsf{vk},m,\sigma)
\,:\,
\begin{array}{l}
\mathsf{crs}\leftarrow\mathsf{Setup}(1^\lambda)\\
(\mathsf{vk},\mathsf{sigk})\leftarrow\mathsf{Gen}(\mathsf{crs})\\
m\leftarrow M(\mathsf{crs},\mathsf{vk})\\
\sigma\leftarrow\mathsf{Sign}(\mathsf{crs},\mathsf{sigk},m)
\end{array}
\right]\ge1-\mathsf{negl}(\lambda).
\end{align}

\paragraph{Security.}
For any QPT adversary $\mathcal{A}$,
\begin{align}
\Pr\left[
\begin{array}{l}
\top\leftarrow\mathsf{Ver}(\mathsf{crs},\mathsf{vk},m_0,\sigma_0)\\
\wedge\,
\top\leftarrow\mathsf{Ver}(\mathsf{crs},\mathsf{vk},m_1,\sigma_1)\\
\wedge\,
m_0\neq m_1
\end{array}
:
\begin{array}{l}
\mathsf{crs}\leftarrow\mathsf{Setup}(1^\lambda)\\
(\mathsf{vk},m_0,\sigma_0,m_1,\sigma_1)\leftarrow\mathcal{A}(1^\lambda,\mathsf{crs})
\end{array}
\right]
\le\mathsf{negl}(\lambda).
\end{align}
\end{definition}

\begin{remark}
\label{rem:OSS}
We make several remarks on the definition of one-shot signatures.

First, we require that $\mathsf{Ver}$ is a classical deterministic polynomial-time algorithm.
This is satisfied by the constructions of \cite{C:ShmZha25,EPRINT:HuaVai25}. 
Classical verification is inherent to the motivation of one-shot signatures~\cite{STOC:AGKZ20}, where all communication is classical.
This requirement is necessary for our proof of \cref{thm:OSS_to_QPKE}.

Second, we require the correctness for adversarially chosen messages: in the correctness of \cref{def:OSS}, the message is chosen by $M$ after $\mathsf{crs}$ and $\mathsf{vk}$ are sampled.
The correctness of one-shot signatures is usually stated for a fixed message, where the probability is taken over $\mathsf{crs}$ and the randomness of $\mathsf{Gen}$ and $\mathsf{Sign}$ after the message is fixed~\cite{STOC:AGKZ20,C:ShmZha25}.
We require this form of correctness because
our proof of \cref{thm:OSS_to_QPKE} applies the correctness to messages that are generated within the security game and are therefore correlated with $\mathsf{crs}$. 
Note that this formulation is implied by perfect correctness, in which an honestly generated signature passes the verification with probability $1$.
The construction of \cite{EPRINT:HuaVai25} achieves perfect correctness, both relative to a classical oracle and in the plain model, and hence satisfies the correctness of \cref{def:OSS}.

Third, we note that the security of \cref{def:OSS} is the weak form of unforgeability, in which the two forged messages must be distinct.
The known constructions achieve the stronger form, in which two distinct signatures on the same message also count as a forgery~\cite{C:ShmZha25,EPRINT:HuaVai25}. Our proof of \cref{thm:OSS_to_QPKE} applies even with the weaker notion.

Finally, our construction requires one-shot signatures for messages that are
longer than verification keys: on every invocation of $\mathsf{PKGen}$,
a concatenation of three verification keys has to be signed.
Known constructions~\cite{EPRINT:HuaVai25,C:ShmZha25} satisfy this property.
\end{remark}

\subsection{QPKE}
\begin{definition}[QPKE with quantum public keys]\label{def:QPKE}
A quantum public-key encryption (QPKE) scheme with quantum public keys consists of four QPT algorithms $(\mathsf{KeyGen},\mathsf{PKGen},\mathsf{Enc},\mathsf{Dec})$ with the following syntax.
\begin{itemize}
\item
$\mathsf{KeyGen}(1^\lambda)\to(\mathsf{vk},\mathsf{sk}_0)$:
The key generation algorithm takes the security parameter $1^\lambda$ as input, and outputs a classical verification key $\mathsf{vk}$ and an initial secret key $\mathsf{sk}_0$, which may be a quantum state.
\item
$\mathsf{PKGen}(\mathsf{sk})\to(\mathsf{pk},\mathsf{sk}')$:
The public key generation algorithm takes a secret key $\mathsf{sk}$ as input, and outputs a quantum public key $\mathsf{pk}$ and an updated secret key $\mathsf{sk}'$, which may be a quantum state.
\item
$\mathsf{Enc}(\mathsf{vk},\mathsf{pk},m)\to\mathsf{ct}$:
The encryption algorithm takes the verification key $\mathsf{vk}$, a quantum public key $\mathsf{pk}$, and a message $m\in\{0,1\}$ as input, 
and outputs a classical ciphertext $\mathsf{ct}$.
\item
$\mathsf{Dec}(\mathsf{sk},\mathsf{ct})\rightarrow m'$:
The decryption algorithm takes the classical part of a secret key $\mathsf{sk}$ and a classical ciphertext $\mathsf{ct}$ as input, and outputs $m'\in\{0,1\}$.
\end{itemize}
Here, a secret key consists of a classical part and a quantum part (the latter may be empty), and $\mathsf{Dec}$ acts only on the classical part.
In particular, secret keys are not consumed by decryption, and polynomially many ciphertexts can be decrypted with the same secret key.

We require the following correctness and security.

\paragraph{Correctness.}
For any polynomial $k$ and any $(m_1,\dots,m_{k(\lambda)})\in\{0,1\}^{k(\lambda)}$,
\begin{align}
\Pr\left[
\begin{array}{c}
\forall i\in[k(\lambda)]\\
m_i\leftarrow\mathsf{Dec}(\mathsf{sk}_{k(\lambda)},\mathsf{ct}_i)
\end{array}
:
\begin{array}{l}
(\mathsf{vk},\mathsf{sk}_0)\leftarrow\mathsf{KeyGen}(1^\lambda)\\
(\mathsf{pk}_j,\mathsf{sk}_j)\leftarrow\mathsf{PKGen}(\mathsf{sk}_{j-1})\ (j=1,\dots,k(\lambda))\\
\mathsf{ct}_i\leftarrow\mathsf{Enc}(\mathsf{vk},\mathsf{pk}_i,m_i)\ (i=1,\dots,k(\lambda))
\end{array}
\right]\ge1-\mathsf{negl}(\lambda),
\end{align}
where the $k(\lambda)$ executions of $\mathsf{Dec}$ take the classical part of $\mathsf{sk}_{k(\lambda)}$ as input, and are therefore well defined.

\paragraph{Security.}
A QPKE scheme $(\mathsf{KeyGen},\mathsf{PKGen},\mathsf{Enc},\mathsf{Dec})$ is secure if for any QPT adversary $\mathcal{A}$,
\begin{align}
\Pr[b'=b]\le\frac{1}{2}+\mathsf{negl}(\lambda)
\end{align}
in the following security game.
\begin{enumerate}
\item
The challenger runs $(\mathsf{vk},\mathsf{sk}_0)\leftarrow\mathsf{KeyGen}(1^\lambda)$ and sends $\mathsf{vk}$ to $\mathcal{A}$.
\item
$\mathcal{A}$ can make public-key queries: 
on the $j$-th query, the challenger runs $(\mathsf{pk}_j,\mathsf{sk}_j)\leftarrow\mathsf{PKGen}(\mathsf{sk}_{j-1})$ and sends $\mathsf{pk}_j$ to $\mathcal{A}$.
\item
$\mathcal{A}$ generates a quantum state over two registers $(\mathbf{P},\mathbf{E})$, sends the register $\mathbf{P}$ to the challenger, and keeps the register $\mathbf{E}$.
\item
The challenger samples $b\leftarrow\{0,1\}$, runs $\mathsf{ct}\leftarrow\mathsf{Enc}(\mathsf{vk},\mathbf{P},b)$, where the state of the register $\mathbf{P}$ is used as the public key, and sends $\mathsf{ct}$ to $\mathcal{A}$.
\item 
$\mathcal{A}$ further makes public-key queries.
\item
$\mathcal{A}$ outputs $b'\in\{0,1\}$.
\end{enumerate}
\end{definition}

\begin{definition}[Everlasting security]\label{def:everlastingQPKE}
A QPKE scheme satisfies everlasting security if the following holds for any adversary $\mathcal{A}$ of the following form: during the 
security game of \cref{def:QPKE}, all operations of $\mathcal{A}$ are QPT; after the interaction with the challenger terminates, $\mathcal{A}$ applies computationally-unbounded operations on all the registers that it keeps, and outputs a bit $b'$. The requirement is that
\begin{align}
\Pr[b'=b]\le\frac{1}{2}+\mathsf{negl}(\lambda).
\end{align}
\end{definition}

\begin{remark}
The security game of \cref{def:QPKE} corresponds to the setting where $\mathsf{vk}$ is distributed over an authenticated classical channel while quantum public keys are distributed over unauthenticated quantum channels: the register $\mathbf{P}$ models the (possibly tampered) quantum public key that the sender actually uses, whereas $\mathsf{vk}$ is guaranteed to be authentic.
\end{remark}

\begin{remark}
Unlike the QPKE of~\cite{C:KMNY24}, our syntax allows $\mathsf{PKGen}$ to be stateful (the secret key is updated and carried over across invocations), and secret keys may be quantum, while $\mathsf{vk}$ and ciphertexts remain classical.
In our construction, the quantum part of the secret key is a single one-shot signing key, and its size does not grow with the number of issued public keys; all other components of the secret key are classical, and $\mathsf{Dec}$ takes only the classical part as input.
A quantum public key is consumed by a single execution of $\mathsf{Enc}$, and therefore each $\mathsf{pk}_i$ is used to encrypt a single message; multi-bit messages can be encrypted bit by bit under independently generated public keys.
\end{remark}
\section{2QKE from Digital Signatures}

\begin{theorem}\label{thm:everlasting-ke}
If one-time secure digital signatures with quantum signatures exist, then 2QKE with everlasting security exist.
\end{theorem}

\begin{proof}
Let $(\KeyGen,\Sign,\Vrfy)$ be a one-time secure digital signature scheme with quantum signatures.
By \cref{rem:pure-signature}, we assume that signatures $|\sigma(0)\rangle$ and $|\sigma(1)\rangle$ are pure states.
We assume without loss of generality that $\sigk$ consists of two registers $(\sigk_0,\sigk_1)$ such that $\Sign(\sigk,m)$ acts only on $\sigk_m$, so that Alice can generate both $|\sigma(0)\rangle$ and $|\sigma(1)\rangle$. Note that this can be satisfied by running any signature algorithm twice, and using the first signing key to generate $|\sigma(0)\rangle$ and the second to generate $|\sigma(1)\rangle$.

Write the verification algorithm in the standard form: $\mathsf{Vrfy}(\vk,m,\cdot)$ appends an ancilla register $\regB'$ initialized to $|0...0\rangle$, applies a unitary $U_{\vk,m}$, and measures a flag qubit in the computational basis, accepting if and only if the outcome is $\top$. Define the projection $\Pi_{\vk,m}$ on the registers $(\regB,\regB')$ and the POVM element $\Lambda_{\vk,m}$ on the register $\regB$ by
\begin{align}
\Pi_{\vk,m}\coloneqq U_{\vk,m}^\dagger(|\top\rangle\langle\top|\otimes I)U_{\vk,m},
\qquad
\Lambda_{\vk,m}\coloneqq\langle 0...0|_{\regB'}\Pi_{\vk,m}|0...0\rangle_{\regB'},
\label{eq:standard-form}
\end{align}
so that 
$
\Pr[\top\leftarrow\mathsf{Vrfy}(\vk,m,\sigma)]=\mathrm{Tr}[\Lambda_{\vk,m}\sigma] 
$
for every state $\sigma$ of the register $\regB$.

We construct a 2QKE as follows.
\begin{enumerate}
\item
Alice runs $(\vk,\sigk)\gets\KeyGen(1^\secp)$.
\item
Alice runs $|\sigma(0)\rangle\gets\Sign(\sigk,0)$
and $|\sigma(1)\rangle\gets\Sign(\sigk,1)$.
\item
By using the controlled-SWAP gate, Alice generates
\begin{align}
|0\rangle_{\regA}|\sigma(0)\rangle_{\regB}|\sigma(1)\rangle_{\regC}+
|1\rangle_{\regA}|\sigma(1)\rangle_{\regB}|\sigma(0)\rangle_{\regC}.
\end{align}
\item
Alice applies a random Pauli $P$ on $\regC$ to get
\begin{align}
|0\rangle_{\regA}|\sigma(0)\rangle_{\regB}P|\sigma(1)\rangle_{\regC}+
|1\rangle_{\regA}|\sigma(1)\rangle_{\regB}P|\sigma(0)\rangle_{\regC}.
\end{align}
\item 
Alice samples $a\gets\bit$ and applies $Z^a$ on $\regA$ to get
\begin{align}
|0\rangle_{\regA}|\sigma(0)\rangle_{\regB}P|\sigma(1)\rangle_{\regC}+
(-1)^a|1\rangle_{\regA}|\sigma(1)\rangle_{\regB}P|\sigma(0)\rangle_{\regC}.
\end{align}
\item
Alice sends the state to Bob.
Alice also sends $\vk$ to Bob.
\item
Bob appends $|0...0\rangle_{\regB'}$ and measures $\regA\regB\regB'$ with $\{\Pi^\vk,I-\Pi^\vk\}$,
where 
\begin{align}
\Pi^\vk\coloneqq\sum_{m\in\bit}|m\rangle\langle m|_\regA\otimes\Pi_{\vk,m}.
\end{align}
If the result is $I-\Pi^\vk$, he outputs a random bit, and sends the maximally-mixed state to Alice.
If the result is $\Pi^\vk$, the post-measurement state is, due to the gentle measurement lemma,
negligibly close to
\begin{align}
\left(|0\rangle_\regA|\sigma(0)\rangle_\regB P|\sigma(1)\rangle_\regC+
(-1)^{a}|1\rangle_\regA|\sigma(1)\rangle_\regB P|\sigma(0)\rangle_\regC\right)\otimes|0...0\rangle_{\regB'}.
\end{align}
Bob discards $\regB'$.
Bob samples $b\gets\bit$ and applies $Z^b$ on $\regA$
to get
\begin{align}
|0\rangle_\regA|\sigma(0)\rangle_\regB P|\sigma(1)\rangle_\regC+
(-1)^{a\oplus b}|1\rangle_\regA|\sigma(1)\rangle_\regB P|\sigma(0)\rangle_\regC,
\end{align}
and sends it to Alice.
Bob outputs $b$.
\item
Alice applies $P^\dagger$ and the controlled-SWAP to get
$
(|0\rangle+(-1)^{a\oplus b}|1\rangle)|\sigma(0)\rangle|\sigma(1)\rangle.
$
She measures the first qubit in the Hadamard basis, and outputs the measurement result plus $a$.
\end{enumerate}

Correctness is clear.
Let us show everlasting security.
Without loss of generality, Eve works as follows: she reads $\vk$,
prepares her register $\regE$ initialized to $|0...0\rangle$, applies a QPT unitary $V_1^\vk$ on the registers $(\regA,\regB,\regC,\regE)$, and forwards $(\regA,\regB,\regC)$ to Bob; when Bob's message is transmitted, she keeps all the registers $(\regA,\regB,\regC,\regE)$, applies a two-outcome POVM $\{D,I-D\}$ on them, which can be computationally unbounded, and outputs the outcome as $e$.

Because $a$ and $P$ are used only in step 8, we can average Alice's message over them: averaging over $a$ dephases the register $\mathbf{A}$, and averaging over $P$ makes the register $\mathbf{C}$ maximally mixed, and therefore Alice's first phase is identical, from Eve's viewpoint, to the following one.
\begin{enumerate}
\item
Alice runs $(\vk,\sigk)\gets\KeyGen(1^\secp)$.
\item 
Alice chooses $\alpha\gets\bit$.
\item
Alice runs $|\sigma(\alpha)\rangle\gets\Sign(\sigk,\alpha)$.
\item
Alice samples $x\gets\bit^n$.
\item 
Alice sends the state
$
|\alpha\rangle_{\regA}|\sigma(\alpha)\rangle_{\regB}|x\rangle_{\regC}
$
to Bob.
Alice also sends $\vk$ to Bob.
\end{enumerate}
Hence hereafter we consider this Alice's first phase.

By the principle of deferred measurement, we can also assume without loss of generality that $\KeyGen$ has the following form: a unitary is applied to $|0...0\rangle$, some registers are measured in the computational basis to get $\vk$, and the quantum $\sigk$ is a state of some unmeasured registers.
Similarly, we can assume without loss of generality that $\Sign$ has the following form: a unitary is applied to $|0...0\rangle$, some registers are measured in the computational basis to get a result $i$, and the pure quantum signature is a state of some unmeasured registers.
Let $t\coloneqq(\vk,i,x)$. 
For the notational simplicity, hereafter we omit $\vk$ of $\Pi_{\vk,m}$, $\Pi^\vk$, $\Lambda_{\vk,m}$ and $V_1^\vk$.
For each $\alpha\in\bit$ and $t$, 
define the normalized pure state
\begin{align}
|\xi_{\alpha}(t)\rangle    
\coloneqq 
V_1(
|\alpha\rangle_\regA
|\sigma(\alpha)\rangle_\regB
|x\rangle_\regC
|0...0\rangle_\regE),
\end{align}
and the unnormalized state
\begin{align}
|\varphi_{m,\alpha}(t)\rangle\coloneqq\left(\langle m|_{\regA}\otimes\Pi_m\otimes I_{\regC,\regE}\right)
(|\xi_\alpha(t)\rangle\otimes|0...0\rangle_{\regB'})
\label{eq:ev-phix}
\end{align}
for $m\in\bit$. 
We can easily show
\begin{align}
(\Pi\otimes I_{\regC,\regE})
(|\xi_\alpha(t)\rangle|0...0\rangle_{\regB'})=\sum_{m\in\bit}|m\rangle_{\regA}\otimes|\varphi_{m,\alpha}(t)\rangle.
\label{aaa}
\end{align}

For $\beta\in\bit$, let $\sigma_{\beta}$ denote the state of all the registers that Eve keeps at the end of the protocol, 
conditioned on that Bob's output is $\beta$.
Because Bob's output is a uniformly random bit both when Bob's test outputs $\top$ 
and when it outputs $\bot$, Bob's output is independent of $t$, $\alpha$, and the outcome of Bob's test, and hence
\begin{align}
&\sigma_{\beta}=\mathop{\mathbb{E}}_{t,\alpha}\left[(1-p_\alpha(t))\rho_{\mathrm{rej}}(t,\alpha)\right]\\
&+\mathbb{E}_{t,\alpha}\left[
\mbox{Tr}_{\regB'}
(Z^{\beta}_\regA\otimes I)(\Pi\otimes I_{\regC,\regE})
(|\xi_\alpha(t)\rangle\langle\xi_\alpha(t)|\otimes|0...0\rangle\langle0...0|_{\regB'})
(\Pi\otimes I_{\regC,\regE})(Z^{\beta}_\regA\otimes I)\right],\label{eq:ev-residual}
\end{align}
where $p_\alpha(t)$ is the probability that Bob's test outputs $\top$,
and $\rho_{\mathrm{rej}}(t,\alpha)$ is the state of the registers that Eve keeps 
conditioned on $(t,\alpha)$ and on Bob's test outputting $\bot$, which is independent of $\beta$. 
Substituting \cref{aaa} into \cref{eq:ev-residual}, we obtain
\begin{align}
\sigma_1-\sigma_0
=-\sum_{\alpha\in\bit}\mathbb{E}_t\left[
\mbox{Tr}_{\regB'}
\left[
|0\rangle\langle1|_{\regA}\otimes|\varphi_{0,\alpha}(t)\rangle\langle\varphi_{1,\alpha}(t)|
+|1\rangle\langle0|_{\regA}\otimes|\varphi_{1,\alpha}(t)\rangle\langle\varphi_{0,\alpha}(t)|
\right]
\right].\label{eq:ev-difference}
\end{align}

For any POVM element $0\preceq D\preceq I$,
\begin{align}
\Pr[e=\beta]
=\frac{1}{2}\Tr[(I-D)\sigma_0]+\frac{1}{2}\Tr[D\sigma_1]
=\frac{1}{2}+\frac{1}{2}\Tr\left[D(\sigma_1-\sigma_0)\right]
\leq\frac{1}{2}+\frac{1}{4}\norm{\sigma_1-\sigma_0}_1.
\end{align}
By \cref{eq:ev-difference}, the triangle inequality, $\norm{|0\rangle\langle1|_{\regA}\otimes M}_1=\norm{M}_1$, and $\norm{M^\dagger}_1=\norm{M}_1$,
\begin{align}
\norm{\sigma_1-\sigma_0}_1
\leq2\sum_{\alpha\in\bit}\norm{\mathop{\mathbb{E}}_t\left[|\varphi_{0,\alpha}(t)\rangle\langle\varphi_{1,\alpha}(t)|\right]}_1.
\end{align}
It therefore suffices to show, for each $\alpha\in\bit$,
\begin{align}
\norm{\mathop{\mathbb{E}}_t\left[|\varphi_{0,\alpha}(t)\rangle\langle\varphi_{1,\alpha}(t)|\right]}_1\leq\negl(\secp).\label{eq:ev-goal}
\end{align}

Define
\begin{align}
\eta_\alpha\coloneqq\mathop{\mathbb{E}}_{t}\left[\norm{|\varphi_{1-\alpha,\alpha}(t)\rangle}^2\right].
\label{bbb}
\end{align}
Hence, by the triangle inequality, $\norm{|u\rangle\langle v|}_1=\norm{|u\rangle}\cdot\norm{|v\rangle}$, $\norm{|\varphi_{\alpha,\alpha}\rangle}\leq1$, and Jensen's inequality,
\begin{align}
\norm{\mathop{\mathbb{E}}_t\left[|\varphi_{0,\alpha}(t)\rangle\langle\varphi_{1,\alpha}(t)|\right]}_1
\leq\mathop{\mathbb{E}}_{t}\left[\norm{|\varphi_{0,\alpha}(t)\rangle}\cdot\norm{|\varphi_{1,\alpha}(t)\rangle}\right]
\leq\mathop{\mathbb{E}}_{t}\left[\norm{|\varphi_{1-\alpha,\alpha}(t)\rangle}\right]
\leq\sqrt{\eta_\alpha}.\label{eq:ev-chain}
\end{align}
\Cref{eq:ev-goal} then follows from the following claim.

\begin{claim}\label{clm:eta-negl}
For each $\alpha\in\{0,1\}$, $\eta_\alpha\leq\mathsf{negl}(\lambda)$.
\end{claim}

\begin{proof}
Fix $\alpha\in\{0,1\}$. 
Because $|m\rangle\langle m|_\regA\otimes\Pi_m\otimes I_{\regC,\regE}$ is a projection, and because $\langle 0...0|_{\regB'}\Pi_m|0...0\rangle_{\regB'}=\Lambda_m$ by \cref{eq:standard-form}, we have, for each $m\in\{0,1\}$ and conditioned on any fixed $t$,
\begin{align}
\||\varphi_{m,\alpha}(t)\rangle\|^2
&=(\langle\xi_\alpha(t)|\otimes\langle 0...0|_{\regB'})\left(|m\rangle\langle m|_\regA\otimes\Pi_m\otimes I_{\regC,\regE}\right)(|\xi_\alpha(t)\rangle\otimes|0...0\rangle_{\regB'})
\label{eq:phi-projection}\\
&=\langle\xi_\alpha(t)|\left(|m\rangle\langle m|_\regA\otimes\Lambda_m\otimes I_{\regC,\regE}\right)|\xi_\alpha(t)\rangle,
\label{eq:phi-lambda}
\end{align}
where $\Lambda_m$ acts on the register $\regB$. In particular,
\begin{align}
\eta_\alpha=\mathbb{E}_{t}\left[\langle\xi_\alpha(t)|\left(|1-\alpha\rangle\langle 1-\alpha|_\regA\otimes\Lambda_{1-\alpha}\otimes I_{\regC,\regE}\right)|\xi_\alpha(t)\rangle\right].
\label{eq:eta-lambda}
\end{align}

For each $\alpha\in\bit$, consider the following QPT adversary $G^\alpha=(G_1^\alpha,G_2^\alpha)$ against the one-time security of the digital signature scheme.
\begin{itemize}
\item $G_1^\alpha(\vk)$: Output the message $m\coloneqq\alpha$ and $st\coloneqq \vk$.
\item $G_2^\alpha(\sigma,st)$: Sample $x\leftarrow\{0,1\}^n$ and prepare $|\alpha\rangle_\regA\otimes\sigma_\regB\otimes|x\rangle_\regC\otimes|0...0\rangle_\regE$, where $\sigma$ is the signature given by the challenger. Apply $V_1$ on $(\regA,\regB,\regC,\regE)$, and measure the register $\regA$ in the computational basis. If the outcome is not $1-\alpha$, abort. Otherwise, output the state of the register $\regB$ as $\sigma'$.
\end{itemize}
We show that the probability that $G_2^\alpha$ does not abort and the challenger's verification accepts is exactly $\eta_\alpha$. In the one-time security experiment against $G^\alpha$, the challenger runs $(\vk,\sigk)\leftarrow\mathsf{KeyGen}(1^\lambda)$ and $\sigma\leftarrow\mathsf{Sign}(\sigk,\alpha)$. By the normal form of $\mathsf{Sign}$ and \cref{rem:pure-signature}, conditioned on $\vk$ and the result $i$ of the internal measurement of this signing, $\sigma$ is the pure state $|\sigma(\alpha)\rangle$. 
Condition on $t=(\vk,i,x)$. After $G_2^\alpha$ applies $V_1$, the state of the registers $(\regA,\regB,\regC,\regE)$ is $|\xi_\alpha(t)\rangle$. 
Define 
\begin{align}
|\chi\rangle\coloneqq(\langle 1-\alpha|_\regA\otimes I_{\regB,\regC,\regE})|\xi_\alpha(t)\rangle.
\end{align}
The probability that the measurement of $\regA$ yields $1-\alpha$ is $\||\chi\rangle\|^2$, and, conditioned on this outcome, the challenger's verification $\mathsf{Vrfy}(\vk,1-m,\cdot)=\mathsf{Vrfy}(\vk,1-\alpha,\cdot)$ accepts with probability $\langle\chi|(\Lambda_{1-\alpha}\otimes I_{\regC,\regE})|\chi\rangle/\||\chi\rangle\|^2$, because its acceptance probability is the expectation of $\Lambda_{1-\alpha}$ in the reduced state of the register $\regB$. The probability that $G_2^\alpha$ does not abort and the challenger accepts is therefore
\begin{align}
\langle\chi|(\Lambda_{1-\alpha}\otimes I_{\regC,\regE})|\chi\rangle
=\langle\xi_\alpha(t)|\left(|1-\alpha\rangle\langle 1-\alpha|_\regA\otimes\Lambda_{1-\alpha}\otimes I_{\regC,\regE}\right)|\xi_\alpha(t)\rangle
=\||\varphi_{1-\alpha,\alpha}(t)\rangle\|^2,
\label{eq:win-prob}
\end{align}
where the last equality is \cref{eq:phi-lambda}. 
Averaging over $t=(\vk,i,x)$, $G^\alpha$ wins the one-time security experiment with probability $\eta_\alpha$ by \cref{bbb}, where aborting counts as losing. The one-time security of the digital signature scheme therefore gives $\eta_\alpha\leq\mathsf{negl}(\lambda)$.
\end{proof}

\end{proof}

\section{EV-StatePuzzs from 2QKE}

\begin{theorem}
\label{lem:2QKE-to-gap}
If 2QKE exists, 
then $\frac{1}{2}$-sound EV-StatePuzzs exist. 
\end{theorem}

\begin{proof}
Assume that a 2QKE exists. We can assume without loss of generality that Alice's first phase has the following form: she applies a unitary $U$ on $|0...0\rangle$ over three registers $(\regS,\regM,\regW)$ to generate the state
\begin{align}
U|0...0\rangle=\sum_s C_s|s\rangle_\regS|\Phi_s\rangle_{\regM,\regW},
\end{align}
where each $C_s\geq0$, $\sum_s C_s^2=1$, and each $|\Phi_s\rangle$ is a normalized pure state; she measures $\regS$ in the computational basis to obtain $s$; she sends $s$ over the classical channel and the register $\regM$ over the quantum channel to Bob, and keeps the register $\regW$. Alice's second phase is a quantum algorithm that takes $s$, the register $\regW$, and the quantum state returned from Bob as input, and outputs the bit $\alpha$.

We construct an EV-StatePuzz $(\mathsf{Samp},\mathsf{Ver})$ as follows.
\begin{itemize}
\item $\mathsf{Samp}(1^\lambda)$: Apply $U$ on $|0...0\rangle$ and measure $\regS$ in the computational basis to obtain $s$. Output the puzzle $s$ and the answer $\psi_s\coloneqq|\Phi_s\rangle$, i.e., the state of the registers $(\regM,\regW)$.
\item $\mathsf{Ver}(1^\lambda,s,\rho)$: The input $\rho$ is a state of the registers $(\regM,\regW)$. Run Bob's honest algorithm on $s$ and the register $\regM$; it outputs a bit $\beta$ and returns a quantum state. Run Alice's second phase on $s$, the register $\regW$, and the returned state; it outputs a bit $\alpha$. Output $\top$ if and only if $\alpha=\beta$.
\end{itemize}
Both algorithms are QPT.

Correctness is clear. 
Let us show security. Let $\cA$ be a QPT adversary and write
\begin{align}
p(\lambda)\coloneqq\Pr[\top\leftarrow\mathsf{Ver}(1^\lambda,s,\cA(1^\lambda,s)):(s,\psi_s)\leftarrow\mathsf{Samp}(1^\lambda)].
\end{align}
We construct the following QPT Eve:
\begin{enumerate}
\item Eve intercepts and discards the quantum message on the register $\regM$. She reads $s$ from the classical channel without modifying it.
\item She runs $\cA(1^\lambda,s)$ to obtain a state of fresh registers $(\regM',\regW')$. She sends the register $\regM'$ to Bob over the quantum channel, and keeps the register $\regW'$.
\item Bob runs his honest algorithm on $s$ and the received register $\regM'$, outputs $\beta$, and returns a quantum state to Alice. Eve intercepts the returned state. (She sends nothing to Alice.)
\item Eve runs Alice's second phase on $s$, the register $\regW'$, and the intercepted state, and outputs its output as $e$.
\end{enumerate}
Conditioned on any fixed $s$, the state of the registers $(\regM',\regW')$ and the process applied to it are identical in the attacked execution and in the experiment of $\mathsf{Ver}(1^\lambda,s,\cA(1^\lambda,s))$, where the roles of $(\beta,\alpha)$ in the latter are played by $(\beta,e)$ in the former; moreover, the distribution of $s$ is the same in the two experiments. Therefore
\begin{align}
\Pr[e=\beta]=p(\lambda).
\end{align}
The security of the 2QKE gives $p(\lambda)\leq\frac{1}{2}+\mathsf{negl}(\lambda)$, which is the claimed soundness.
\end{proof}


\begin{lemma}
Let $p$ be a polynomial and let $\delta:N\to[0,1]$ satisfying $\delta(\secp)\le1-1/p(\secp)$.
If $\delta$-sound EV-StatePuzzs exist, then EV-StatePuzzs exist.
\end{lemma}
\begin{proof}
Let $(\mathsf{Samp},\mathsf{Ver})$ be a $\delta$-sound EV-StatePuzz, and let
$(\mathsf{Samp}',\mathsf{Ver}')$ be its $k$-fold parallel repetition with
$k\coloneqq\lambda p(\lambda)$: $\mathsf{Samp}'$ runs $\mathsf{Samp}(1^\lambda)$
independently $k$ times, and $\mathsf{Ver}'$ accepts if and only if all $k$
answer registers are accepted. Correctness follows from the union bound.
The security experiment of $(\mathsf{Samp},\mathsf{Ver})$ is a two-message
quantum interactive protocol between a QPT challenger and the adversary,
and that of $(\mathsf{Samp}',\mathsf{Ver}')$ is exactly its $k$-fold parallel
repetition. Hence, by the parallel repetition theorem for three-message
computationally-secure quantum interactive protocols with efficient
challengers~\cite{STOC:BQSY24}, any QPT adversary wins the latter with
probability at most $(\delta(\lambda)+{\rm negl}(\lambda))^{k}+{\rm negl}(\lambda)
\le(1-\tfrac{1}{2p(\lambda)})^{\lambda p(\lambda)}+{\rm negl}(\lambda)
\le e^{-\lambda/2}+{\rm negl}(\lambda)\le{\rm negl}(\lambda)$.
\end{proof}

\ifnum\submission=1
\else
\section{QPKE from One-Shot Signatures}

\begin{theorem}
\label{thm:OSS_to_QPKE}
If one-shot signatures exist, then everlasting-secure QPKE with quantum public keys exist.    
\end{theorem}

\begin{proof}
Let $(\mathsf{Setup},\mathsf{Gen},\mathsf{Sign},\mathsf{Ver})$ be a one-shot signature scheme.
Without loss of generality, we assume that verification keys have a fixed polynomial length and that all signatures are $\ell$-bit strings for a polynomial $\ell$, by padding.
For $x,y\in\{0,1\}^{\ell}$, we write $x\cdot y\coloneqq\bigoplus_{i=1}^{\ell}x_iy_i$.
We use two registers $\mathbf{A}$ and $\mathbf{B}$, consisting of one qubit and $\ell$ qubits, respectively.
For a pair $u=(u^0,u^1)$ of verification keys, define the projection
\begin{align}
\Pi^{u}\coloneqq\sum_{\alpha\in\{0,1\}}|\alpha\rangle\langle\alpha|_{\mathbf{A}}\otimes\sum_{\beta\in\{0,1\}^{\ell}:\,\top=\mathsf{Ver}(\mathsf{crs},u^{\alpha},\alpha,\beta)}|\beta\rangle\langle\beta|_{\mathbf{B}}.\label{eq:leafproj}
\end{align}
Because $\mathsf{Ver}$ is classical and deterministic, $\Pi^{u}$ is diagonal in the computational basis, and the projective measurement $\{\Pi^{u},I-\Pi^{u}\}$ can be implemented in QPT by computing $\mathsf{Ver}$ coherently into a flag qubit and measuring the flag.

We construct a QPKE scheme as follows.
\begin{itemize}
\item $\mathsf{KeyGen}(1^\lambda)$: Run $\mathsf{crs}\leftarrow\mathsf{Setup}(1^\lambda)$ and $(w_1,\mathsf{sigk}_1)\leftarrow\mathsf{Gen}(\mathsf{crs})$. Output $\mathsf{vk}\coloneqq(\mathsf{crs},w_1)$ and $\mathsf{sk}_0\coloneqq(\mathsf{crs},w_1,\mathsf{sigk}_1)$ together with a counter initialized to $0$ and an empty table. Only $\mathsf{sigk}_1$ is quantum.
\item $\mathsf{PKGen}(\mathsf{sk}_{t-1})$ (the $t$-th invocation; $\mathsf{sk}_{t-1}$ contains $\mathsf{crs}$, the classical data $(\mu_s,\tau_s,a_s,\sigma_s(0),\sigma_s(1))_{s\le t-1}$ defined below, and the signing key $\mathsf{sigk}_t$):
\begin{enumerate}
\item Run $(u_t^0,\mathsf{sigk}_t^0)\leftarrow\mathsf{Gen}(\mathsf{crs})$, $(u_t^1,\mathsf{sigk}_t^1)\leftarrow\mathsf{Gen}(\mathsf{crs})$, and $(w_{t+1},\mathsf{sigk}_{t+1})\leftarrow\mathsf{Gen}(\mathsf{crs})$. We call $(u_t^0,u_t^1)$ the $t$-th leaf keys.
\item Set $\mu_t\coloneqq u_t^0\|u_t^1\|w_{t+1}$ and run $\tau_t\leftarrow\mathsf{Sign}(\mathsf{crs},\mathsf{sigk}_t,\mu_t)$. This consumes $\mathsf{sigk}_t$.
\item For each $\alpha\in\{0,1\}$, run $\sigma_t(\alpha)\leftarrow\mathsf{Sign}(\mathsf{crs},\mathsf{sigk}_t^{\alpha},\alpha)$. This consumes $\mathsf{sigk}_t^0$ and $\mathsf{sigk}_t^1$.
\item Sample $a_t\leftarrow\{0,1\}$ and prepare the state
\begin{align}
\frac{1}{\sqrt{2}}\left(|0\rangle_{\mathbf{A}}|\sigma_t(0)\rangle_{\mathbf{B}}+(-1)^{a_t}|1\rangle_{\mathbf{A}}|\sigma_t(1)\rangle_{\mathbf{B}}\right).\label{eq:pkstate}
\end{align}
\item Output $\mathsf{pk}_t\coloneqq\left(t,(\mu_s,\tau_s)_{s\le t},\text{the state of }(\mathbf{A},\mathbf{B})\right)$ and $\mathsf{sk}_t$, which is $\mathsf{sk}_{t-1}$ with $\mathsf{sigk}_t$ removed and $\mathsf{sigk}_{t+1}$ and $(\mu_t,\tau_t,a_t,\sigma_t(0),\sigma_t(1))$ appended.
\end{enumerate}
\item $\mathsf{Enc}(\mathsf{vk},\mathsf{pk},m)$: Parse $\mathsf{vk}=(\mathsf{crs},w_1)$. Measure the classical part of $\mathsf{pk}$ in the computational basis (this does nothing for honestly generated public keys), and parse the outcome as $(t,(\mu'_s,\tau'_s)_{s\le t})$ and each $\mu'_s$ as $u'^{0}_s\|u'^{1}_s\|w'_{s+1}$. If the parsing fails or $t=0$, output $\mathsf{ct}\coloneqq\bot$.
\begin{enumerate}
\item (Chain check.) Set $w'_1\coloneqq w_1$ and check $\top=\mathsf{Ver}(\mathsf{crs},w'_s,\mu'_s,\tau'_s)$ for all $s\in\{1,\dots,t\}$. If any check fails, output $\mathsf{ct}\coloneqq\bot$.
\item (Leaf check.) Set $u'\coloneqq(u'^0_t,u'^1_t)$ and measure the registers $(\mathbf{A},\mathbf{B})$ of $\mathsf{pk}$ with $\{\Pi^{u'},I-\Pi^{u'}\}$. If the outcome is $I-\Pi^{u'}$, output $\mathsf{ct}\coloneqq\bot$.
\item Apply $Z^m$ on $\mathbf{A}$, measure $(\mathbf{A},\mathbf{B})$ in the Hadamard basis to obtain $(c,d)\in\{0,1\}\times\{0,1\}^{\ell}$, and output $\mathsf{ct}\coloneqq(t,c,d)$.
\end{enumerate}
\item $\mathsf{Dec}(\mathsf{sk},\mathsf{ct})$: If $\mathsf{ct}=\bot$, output a uniformly random bit. Otherwise, parse $\mathsf{ct}=(t,c,d)$, retrieve $(a_t,\sigma_t(0),\sigma_t(1))$ from $\mathsf{sk}$, and output $m'\coloneqq c\oplus a_t\oplus d\cdot(\sigma_t(0)\oplus\sigma_t(1))$.
\end{itemize}
All algorithms are QPT. Ciphertexts are classical, and $\mathsf{Dec}$ uses only the classical part of the secret key.
The quantum part of $\mathsf{sk}_t$ is the single unconsumed signing key $\mathsf{sigk}_{t+1}$, and hence its size does not grow with the number $t$ of issued public keys; the leaf signing keys $\mathsf{sigk}_t^0,\mathsf{sigk}_t^1$ are consumed within the $t$-th invocation of $\mathsf{PKGen}$, and all other components of $\mathsf{sk}_t$ are classical.

\paragraph{Correctness.}
Consider an honest execution with the $i$-th public key.
By the correctness of the one-shot signature scheme and the union bound, with probability $1-\mathsf{negl}(\lambda)$, all the pairs $(\mu_s,\tau_s)$ for $s\le i$ pass the verification under $w_s$, and $\top=\mathsf{Ver}(\mathsf{crs},u_i^{\alpha},\alpha,\sigma_i(\alpha))$ for both $\alpha\in\{0,1\}$.
Condition on this event.
Then the chain check passes, and the state \cref{eq:pkstate} lies in the image of $\Pi^{u_i}$, so that the leaf check accepts with certainty without changing the state.
After $Z^m$ is applied on $\mathbf{A}$, the state is $\frac{1}{\sqrt{2}}(|0\rangle_{\mathbf{A}}|\sigma_i(0)\rangle_{\mathbf{B}}+(-1)^{a_i\oplus m}|1\rangle_{\mathbf{A}}|\sigma_i(1)\rangle_{\mathbf{B}})$, and the Hadamard-basis measurement returns $(c,d)$ with probability
\begin{align}
\frac{1}{2^{\ell+2}}\left|1+(-1)^{a_i\oplus m\oplus c\oplus d\cdot(\sigma_i(0)\oplus\sigma_i(1))}\right|^2,\label{eq:hadprob}
\end{align}
which vanishes unless $c\oplus a_i\oplus d\cdot(\sigma_i(0)\oplus\sigma_i(1))=m$. Hence $\mathsf{Dec}$ outputs $m$.
Because $(a_i,\sigma_i(0),\sigma_i(1))$ is retained in the classical part of the secret key, this holds for $\mathsf{Dec}(\mathsf{sk}_k,\cdot)$ for any $i\le k$, and the correctness follows from the union bound over the $k$ ciphertexts.

\paragraph{Everlasting security.}
Let $\mathcal{A}$ be an adversary as in \cref{def:everlastingQPKE}: all operations of $\mathcal{A}$ are QPT during the interaction with the challenger, and $\mathcal{A}$ applies computationally-unbounded operations afterwards.
Let $q'$ be the number of public-key queries made before the challenge.
Without loss of generality, $\mathcal{A}$ is unitary during the QPT phase, and we write $\mathbf{E}$ for the register of $\mathcal{A}$.

\paragraph{\it Removing the pads.}
For any state $\rho$ of $(\mathbf{A},\mathbf{B})$,
\begin{align}
\frac{1}{2}\sum_{a\in\{0,1\}}(Z^{a}_{\mathbf{A}}\otimes I_{\mathbf{B}})\rho(Z^{a}_{\mathbf{A}}\otimes I_{\mathbf{B}})=\sum_{\alpha\in\{0,1\}}(|\alpha\rangle\langle\alpha|_{\mathbf{A}}\otimes I_{\mathbf{B}})\rho(|\alpha\rangle\langle\alpha|_{\mathbf{A}}\otimes I_{\mathbf{B}}),\label{eq:dephase}
\end{align}
i.e., averaging over the pad is the completely dephasing channel on $\mathbf{A}$.
Applied to the state \cref{eq:pkstate}, it gives $\frac{1}{2}\sum_{\alpha\in\{0,1\}}|\alpha\rangle\langle\alpha|_{\mathbf{A}}\otimes|\sigma_t(\alpha)\rangle\langle\sigma_t(\alpha)|_{\mathbf{B}}$.
The pad $a_t$ is used nowhere else in the security game, because $\mathsf{Dec}$ is never executed in the security game of \cref{def:QPKE}.
Therefore, the state of all registers ever given to $\mathcal{A}$ is unchanged if step 4 of $\mathsf{PKGen}$ is replaced with the following: sample $\alpha_t\leftarrow\{0,1\}$ and set the state of $(\mathbf{A},\mathbf{B})$ to $|\alpha_t\rangle_{\mathbf{A}}|\sigma_t(\alpha_t)\rangle_{\mathbf{B}}$.
In this game, $\sigma_t(1-\alpha_t)$ is never used.
Since the signing $\sigma_t(1-\alpha_t)\leftarrow\mathsf{Sign}(\mathsf{crs},\mathsf{sigk}_t^{1-\alpha_t},1-\alpha_t)$ acts only on registers held by the challenger and its output is never sent to $\mathcal{A}$, the state of all registers ever given to $\mathcal{A}$ is further unchanged if the challenger omits this signing and keeps $\mathsf{sigk}_t^{1-\alpha_t}$ unconsumed.
Hence the security game is statistically identical to the following modified security game: on the $t$-th public-key query, the challenger runs steps 1 and 2 of $\mathsf{PKGen}$, samples $\alpha_t\leftarrow\{0,1\}$, runs $\sigma_t\leftarrow\mathsf{Sign}(\mathsf{crs},\mathsf{sigk}_t^{\alpha_t},\alpha_t)$, keeps $\mathsf{sigk}_t^{1-\alpha_t}$ unconsumed, and sends
\begin{align}
\mathsf{pk}_t\coloneqq\left(t,(\mu_s,\tau_s)_{s\le t},|\alpha_t\rangle_{\mathbf{A}}|\sigma_t\rangle_{\mathbf{B}}\right).\label{eq:modpk}
\end{align}
We stress that this replacement is an equality of the state of the registers of $\mathcal{A}$, and is therefore valid even against the unbounded phase of $\mathcal{A}$.
Hereafter we consider the modified security game.

\paragraph{\it The challenge.}
Let $T$ denote the entire classical transcript of the modified security game up to and including the measurement of the classical part of the public key performed by $\mathsf{Enc}$ at the challenge, namely $\mathsf{crs}$, $w_1$, $(\mu_s,\tau_s,\alpha_s,\sigma_s)_{s\le q'}$, and the outcome $(t',(\mu'_s,\tau'_s)_{s\le t'})$ of the measurement.
The distribution of $T$ does not depend on $b$.
Every input to $\mathcal{A}$ before the challenge is a classical string contained in $T$, and $\mathcal{A}$ is unitary.
Hence, conditioned on $T$, the joint state of the registers $(\mathbf{A},\mathbf{B})$ of the submitted public key and $\mathbf{E}$ is a normalized pure state, which we denote by $|\xi(T)\rangle$, and the registers held by the challenger, namely the unconsumed signing keys, are in a state that is in tensor product with $|\xi(T)\rangle$ and does not depend on $b$.

Let $\sigma_b(T)$ be the state of $(\mathbf{A},\mathbf{B},\mathbf{E})$ immediately after $\mathsf{Enc}$ has performed the leaf check and, on acceptance, applied $Z^b$ on $\mathbf{A}$, conditioned on $T$ and on the challenge bit being $b$; if $t'=0$ or the chain check fails, we set $\sigma_b(T)\coloneqq|\xi(T)\rangle\langle\xi(T)|$.
Everything that happens afterwards, namely the Hadamard-basis measurement of $(\mathbf{A},\mathbf{B})$, the delivery of $\mathsf{ct}$, the answers to further public-key queries, and the QPT operations of $\mathcal{A}$, is a quantum channel that may depend on $T$ but not on $b$, applied to $\sigma_b(T)$ and to the registers of the challenger.
Let $\omega_b$ be the state of the registers of $\mathcal{A}$ at the end of the QPT phase conditioned on $b$.
By the above, $\omega_b=\sum_T\Pr[T]\,\Phi_T(\sigma_b(T))$ for some quantum channels $\Phi_T$ that do not depend on $b$, and hence, by the monotonicity of the trace distance,
\begin{align}
\|\omega_1-\omega_0\|_1\le\sum_T\Pr[T]\,\|\sigma_1(T)-\sigma_0(T)\|_1.\label{eq:cond}
\end{align}
The unbounded phase of $\mathcal{A}$ is a two-outcome POVM $\{D,I-D\}$ on $\omega_b$, and therefore
\begin{align}
\Pr[b'=b]=\frac{1}{2}\mathrm{Tr}[(I-D)\omega_0]+\frac{1}{2}\mathrm{Tr}[D\omega_1]=\frac{1}{2}+\frac{1}{2}\mathrm{Tr}[D(\omega_1-\omega_0)]\le\frac{1}{2}+\frac{1}{4}\|\omega_1-\omega_0\|_1.\label{eq:pov}
\end{align}
It thus suffices to bound the right-hand side of \cref{eq:cond}.

Fix $T$ such that $t'\ge1$ and the chain check passes, and write $u'\coloneqq(u'^0_{t'},u'^1_{t'})$.
For $\alpha\in\{0,1\}$, define the unnormalized vector on $(\mathbf{B},\mathbf{E})$
\begin{align}
|\varphi_\alpha(T)\rangle\coloneqq(\langle\alpha|_{\mathbf{A}}\otimes I_{\mathbf{B},\mathbf{E}})(\Pi^{u'}\otimes I_{\mathbf{E}})|\xi(T)\rangle,\label{eq:phi}
\end{align}
so that $(\Pi^{u'}\otimes I_{\mathbf{E}})|\xi(T)\rangle=\sum_{\alpha\in\{0,1\}}|\alpha\rangle_{\mathbf{A}}\otimes|\varphi_\alpha(T)\rangle$ and $\||\varphi_\alpha(T)\rangle\|\le1$.
The branch on which the leaf check rejects does not depend on $b$, and the unnormalized state on the accepting branch after $Z^b$ is applied is $\sum_{\alpha\in\{0,1\}}(-1)^{b\alpha}|\alpha\rangle_{\mathbf{A}}\otimes|\varphi_\alpha(T)\rangle$.
Hence
\begin{align}
\sigma_1(T)-\sigma_0(T)=-2\left(|0\rangle\langle1|_{\mathbf{A}}\otimes|\varphi_0(T)\rangle\langle\varphi_1(T)|+|1\rangle\langle0|_{\mathbf{A}}\otimes|\varphi_1(T)\rangle\langle\varphi_0(T)|\right),\label{eq:diff}
\end{align}
and, by the triangle inequality and $\||u\rangle\langle v|\|_1=\||u\rangle\|\cdot\||v\rangle\|$,
\begin{align}
\|\sigma_1(T)-\sigma_0(T)\|_1\le4\,\||\varphi_0(T)\rangle\|\cdot\||\varphi_1(T)\rangle\|.\label{eq:pertranscript}
\end{align}

We call $T$ good if $1\le t'\le q'$, the chain check passes, and $\mu'_s=\mu_s$ for all $s\le t'$, i.e., the submitted chain is a prefix of the honest chain.
For good $T$, we have $u'=(u^0_{t'},u^1_{t'})$: the leaf keys of the submitted public key are the honest $t'$-th leaf keys.
We call $T$ bad if $t'\ge1$ and the chain check passes but $T$ is not good, and let $\mathrm{Bad}$ be the event that $T$ is bad.
Since $\sigma_1(T)=\sigma_0(T)$ unless $T$ is good or bad, \cref{eq:pertranscript} and $\||\varphi_\alpha(T)\rangle\|\le1$ give
\begin{align}
\sum_T\Pr[T]\,\|\sigma_1(T)-\sigma_0(T)\|_1
\le4\Pr[\mathrm{Bad}]+4\sum_{T:\,\mathrm{good}}\Pr[T]\,\||\varphi_{1-\alpha_{t'}}(T)\rangle\|
\le4\Pr[\mathrm{Bad}]+4\sqrt{\eta},\label{eq:main}
\end{align}
where
\begin{align}
\eta\coloneqq\sum_{T:\,\mathrm{good}}\Pr[T]\,\||\varphi_{1-\alpha_{t'}}(T)\rangle\|^2,\label{eq:eta}
\end{align}
and the last inequality of \cref{eq:main} is the Cauchy--Schwarz inequality.
The theorem follows from the following two claims.

\begin{claim}
\begin{align}
\Pr[\mathrm{Bad}]\le\mathsf{negl}(\lambda).
\label{clm:bad}
\end{align}
\end{claim}
\begin{proof}
We construct a QPT adversary $\mathcal{B}_1$ against the security of the one-shot signature scheme.
Recall that in the security game of \cref{def:OSS} there is no signing oracle, and the adversary outputs a verification key of its own choice together with valid signatures on two distinct messages under it.
The verification key output by $\mathcal{B}_1$ will be one of the chain keys $w_1,\dots,w_{q'+1}$ generated by $\mathcal{B}_1$ itself in the simulation below; in particular, $\mathcal{B}_1$ does not have to fix a target key in advance.

$\mathcal{B}_1(1^\lambda,\mathsf{crs})$ runs the modified security game by itself, playing the role of the challenger against $\mathcal{A}$, which is run as a subroutine: it runs $(w_1,\mathsf{sigk}_1)\leftarrow\mathsf{Gen}(\mathsf{crs})$, using the given $\mathsf{crs}$ in place of running $\mathsf{Setup}$, sends $\mathsf{vk}\coloneqq(\mathsf{crs},w_1)$ to $\mathcal{A}$, and answers each public-key query of $\mathcal{A}$ as the challenger of the modified security game does.
The simulation is exact, because every operation of the challenger is an invocation of $\mathsf{Gen}$, $\mathsf{Sign}$, or $\mathsf{Ver}$ under $\mathsf{crs}$, or a public operation, and involves no secret unknown to $\mathcal{B}_1$.
As a by-product, when $\mathcal{A}$ submits its public key at the challenge, $\mathcal{B}_1$ holds the honest chain data $(\mu_s,\tau_s)_{s\le q'}$ and the unconsumed signing key $\mathsf{sigk}_{q'+1}$ corresponding to $w_{q'+1}$.
$\mathcal{B}_1$ then measures the classical part of the submitted public key, obtaining $(t',(\mu'_s,\tau'_s)_{s\le t'})$, and halts the simulation.

Suppose that $\mathrm{Bad}$ occurs.
Then $t'\ge1$ and, setting $w'_1\coloneqq w_1$, all the certificates verify, i.e., $\top=\mathsf{Ver}(\mathsf{crs},w'_s,\mu'_s,\tau'_s)$ for all $s\le t'$.
Note that the message at each position determines the verification key of the next position, because $w_{s+1}$ is a component of $\mu_s$ and verification keys have a fixed length.
Exactly one of the following two cases occurs.
\begin{itemize}
\item Case 1: there exists $s\le\min(t',q')$ with $\mu'_s\ne\mu_s$. Let $r$ be the smallest such index. Since $\mu'_s=\mu_s$ for all $s<r$, we have $w'_r=w_r$ (for $r=1$ this is $w'_1=w_1$). Hence $(\mu'_r,\tau'_r)$ verifies under the honest key $w_r$, the honest pair $(\mu_r,\tau_r)$ is held by $\mathcal{B}_1$ because $r\le q'$, and $\mu_r\ne\mu'_r$. $\mathcal{B}_1$ outputs $(w_r,\mu_r,\tau_r,\mu'_r,\tau'_r)$.
\item Case 2: $\mu'_s=\mu_s$ for all $s\le\min(t',q')$. Since $T$ is not good, we must have $t'>q'$. In particular, the position $q'+1$ exists, and its verification key is the honest one, $w'_{q'+1}=w_{q'+1}$ (for $q'=0$ this is $w'_1=w_1$). Hence $(\mu'_{q'+1},\tau'_{q'+1})$ is a valid message-signature pair under $w_{q'+1}$, whose signing key $\mathsf{sigk}_{q'+1}$ has never signed and is held by $\mathcal{B}_1$. $\mathcal{B}_1$ picks an arbitrary $\mu''\ne\mu'_{q'+1}$, runs $\tau''\leftarrow\mathsf{Sign}(\mathsf{crs},\mathsf{sigk}_{q'+1},\mu'')$, and outputs $(w_{q'+1},\mu'_{q'+1},\tau'_{q'+1},\mu'',\tau'')$.
\end{itemize}
In both cases, conditioned on $\mathrm{Bad}$, the output of $\mathcal{B}_1$ consists of a verification key and signatures on two distinct messages under it, and the pair coming from the submitted chain verifies because the chain check passed.
The other pair, $(\mu_r,\tau_r)$ in Case 1 and $(\mu'',\tau'')$ in Case 2, verifies with probability $1-\mathsf{negl}(\lambda)$ by the correctness of the one-shot signature scheme and the union bound over the at most $q'+1$ pairs involved; here we use the correctness for adversarially chosen messages, because these messages are correlated with $\mathsf{crs}$ and the verification key.
Hence $\mathcal{B}_1$ wins the security game of \cref{def:OSS} with probability at least $\Pr[\mathrm{Bad}]-\mathsf{negl}(\lambda)$, and the security of the one-shot signature scheme gives $\Pr[\mathrm{Bad}]\le\mathsf{negl}(\lambda)$.
\end{proof}

\begin{claim}
\begin{align}
\eta\le\mathsf{negl}(\lambda).
\label{clm:eta}
\end{align}
\end{claim}
\begin{proof}
We construct a QPT adversary $\mathcal{B}_2$ against the security of the one-shot signature scheme.
The verification key output by $\mathcal{B}_2$ will be one of the leaf keys generated by $\mathcal{B}_2$ itself in the simulation below.

$\mathcal{B}_2(1^\lambda,\mathsf{crs})$ runs the modified security game by itself, playing the role of the challenger against $\mathcal{A}$, which is run as a subroutine, exactly as $\mathcal{B}_1$ does.
As a by-product, when $\mathcal{A}$ submits its public key at the challenge, $\mathcal{B}_2$ holds, for each $t\le q'$, the bit $\alpha_t$ and the unconsumed signing key $\mathsf{sigk}_t^{1-\alpha_t}$ corresponding to the leaf key $u_t^{1-\alpha_t}$.
Here we crucially use the modified security game, in which the challenger signs only with $\mathsf{sigk}_t^{\alpha_t}$.

$\mathcal{B}_2$ measures the classical part of the submitted public key, obtaining $(t',(\mu'_s,\tau'_s)_{s\le t'})$, and checks whether $T$ is good; this is possible because the chain check is a public computation and $\mathcal{B}_2$ has generated $(\mu_s)_{s\le q'}$ itself.
If $T$ is not good, $\mathcal{B}_2$ aborts; this loses nothing, because $\eta$ is a sum over good transcripts.
Otherwise, let $j\coloneqq t'$, so that $u'=(u_j^0,u_j^1)$; note that $j$ is read off the measured chain, and no guessing is involved.
Then, instead of performing the leaf check and the subsequent steps of $\mathsf{Enc}$, $\mathcal{B}_2$ measures the registers $(\mathbf{A},\mathbf{B})$ in the computational basis to obtain $(\alpha^*,\beta^*)$.
If $\alpha^*=1-\alpha_j$ and $\top=\mathsf{Ver}(\mathsf{crs},u_j^{1-\alpha_j},1-\alpha_j,\beta^*)$, then $\mathcal{B}_2$ runs $\tau''\leftarrow\mathsf{Sign}(\mathsf{crs},\mathsf{sigk}_j^{1-\alpha_j},\alpha_j)$ and outputs $(u_j^{1-\alpha_j},1-\alpha_j,\beta^*,\alpha_j,\tau'')$; otherwise it aborts.

We compute the probability that $\mathcal{B}_2$ outputs the tuple.
Condition on a good $T$, which fixes $j$ and $u'=(u_j^0,u_j^1)$.
The state of $(\mathbf{A},\mathbf{B},\mathbf{E})$ at this point is $|\xi(T)\rangle$, and hence, by \cref{eq:leafproj,eq:phi}, the probability that the computational-basis measurement of $(\mathbf{A},\mathbf{B})$ yields $\alpha^*=1-\alpha_j$ together with some $\beta^*$ satisfying $\top=\mathsf{Ver}(\mathsf{crs},u_j^{1-\alpha_j},1-\alpha_j,\beta^*)$ is exactly $\||\varphi_{1-\alpha_j}(T)\rangle\|^2$.
Averaging over the good transcripts, the probability that $\mathcal{B}_2$ outputs the tuple is exactly $\eta$ by \cref{eq:eta}.

Finally, we verify that the output tuple wins the security game of \cref{def:OSS}.
The two messages $1-\alpha_j$ and $\alpha_j$ are distinct.
The pair $(1-\alpha_j,\beta^*)$ passes the verification, because $\mathcal{B}_2$ has checked it and $\mathsf{Ver}$ is classical and deterministic.
The pair $(\alpha_j,\tau'')$ is generated by the honest signing algorithm with the honestly generated signing key $\mathsf{sigk}_j^{1-\alpha_j}$, which has never signed before, and hence passes the verification with probability $1-\mathsf{negl}(\lambda)$ by the correctness of the one-shot signature scheme.
The verification key $u_j^{1-\alpha_j}$ is generated by $\mathcal{B}_2$ itself, which is allowed because in \cref{def:OSS} the adversary outputs the verification key.
Hence $\mathcal{B}_2$, which is QPT, wins the security game of \cref{def:OSS} with probability at least $\eta-\mathsf{negl}(\lambda)$, and the security of the one-shot signature scheme gives $\eta\le\mathsf{negl}(\lambda)$.
\end{proof}

By \cref{clm:bad,clm:eta,eq:cond,eq:pov,eq:main}, $\Pr[b'=b]\le\frac{1}{2}+\mathsf{negl}(\lambda)$.
The security of the one-shot signature scheme is invoked only against the QPT reductions $\mathcal{B}_1$ and $\mathcal{B}_2$, which run entirely within the QPT phase of the game; all other steps are equalities of quantum states or bounds on the trace distance, and the unbounded phase of $\mathcal{A}$ is taken into account in \cref{eq:pov}.
This establishes everlasting security.
\end{proof}

\section{Certified Freeze}

\subsection{Definitions}

\begin{definition}[QPKE in the CRS model]
\label{def:QPKEcrs}
A QPKE scheme in the common reference string (CRS) model is a tuple $(\mathsf{Setup},\mathsf{KeyGen},\mathsf{PKGen},\mathsf{Enc},\mathsf{Dec})$ of QPT algorithms, where $\mathsf{Setup}(1^\lambda)\to\mathsf{crs}$ outputs a classical common reference string $\mathsf{crs}$, and $(\mathsf{KeyGen},\mathsf{PKGen},\mathsf{Enc},\mathsf{Dec})$ are as in \cref{def:QPKE} except that they take $\mathsf{crs}$ as an additional input.
The correctness, the security, and the everlasting security are defined in the same way as in \cref{def:QPKE,def:everlastingQPKE} except that $\mathsf{crs}\leftarrow\mathsf{Setup}(1^\lambda)$ is run by a trusted party at the beginning, and $\mathsf{crs}$ is given to the adversary.
For simplicity, we sometimes omit $\mathsf{crs}$ from the inputs of the algorithms.
\end{definition}

\begin{definition}[Certified freeze]
\label{def:freeze}
A QPKE scheme $(\mathsf{Setup},\mathsf{KeyGen},\mathsf{PKGen},\mathsf{Enc},\mathsf{Dec})$ in the CRS model supports certified freeze if there exist the following three additional algorithms.
\begin{itemize}
\item $\mathsf{Freeze}(\mathsf{sk}) \to (\mathsf{cert},\mathsf{sk}^*,\mathsf{count})$: The freezing algorithm is a QPT algorithm that takes a secret key $\mathsf{sk}$ as input, and outputs a classical certificate $\mathsf{cert}$, a post-freeze secret key $\mathsf{sk}^*$, and a count $\mathsf{count}\in\{0,1,2,\dots\}$ for the total number of produced public keys.
\item $\mathsf{FreezeVer}(\mathsf{vk},\mathsf{cert},\mathsf{count}) \to \top/\bot$: The certificate verification algorithm is a classical deterministic polynomial-time algorithm that takes the verification key $\mathsf{vk}$, a certificate $\mathsf{cert}$, and a count $\mathsf{count}$ as input, and outputs $\top$ or $\bot$.
\item $\mathsf{Index}(\mathsf{vk},\mathsf{ct}) \to i/\bot$: The index algorithm is a classical deterministic polynomial-time algorithm
that takes in a ciphertext $\mathsf{ct}$ as well as the verification key $\mathsf{vk}$ and outputs either an index $i\in\{1,2,\dots\}$ or $\bot$.
\end{itemize}
We do not require that $\mathsf{PKGen}$ remains executable on $\mathsf{sk}^*$; indeed, the point of the freeze is to fix the number of public keys that can be used, as formalized by the freeze binding below.

We require the following four properties.

\noindent\textbf{Index validity.} For any polynomial $k$ and any $m\in \{0,1\}$,
\begin{align}
\Pr\left[\mathsf{Index}(\mathsf{vk},\mathsf{ct}_{k(\lambda)}) = k(\lambda)
:
\begin{array}{l}
\mathsf{crs} \leftarrow \mathsf{Setup}(1^\lambda) \\
(\mathsf{vk},\mathsf{sk}_0) \leftarrow \mathsf{KeyGen}(1^\lambda) \\
(\mathsf{pk}_j,\mathsf{sk}_j) \leftarrow \mathsf{PKGen}(\mathsf{sk}_{j-1})\ (j\in[k(\lambda)])\\
\mathsf{ct}_{k(\lambda)} \gets \mathsf{Enc}(\mathsf{vk},\mathsf{pk}_{k(\lambda)},m)
\end{array}
\right] \geq 1 - \mathsf{negl}(\lambda).
\end{align}

\noindent\textbf{Freeze correctness.} For any polynomial $k$,
\begin{align}
\Pr\left[
\begin{array}{@{}l@{}}
\mathsf{FreezeVer}(\mathsf{vk},\mathsf{cert},\mathsf{count}) = \top\\
\wedge\,\mathsf{count} = k(\lambda)
\end{array}
:
\begin{array}{@{}l@{}}
\mathsf{crs} \leftarrow \mathsf{Setup}(1^\lambda) \\
(\mathsf{vk},\mathsf{sk}_0) \leftarrow \mathsf{KeyGen}(1^\lambda) \\
(\mathsf{pk}_j,\mathsf{sk}_j) \leftarrow \mathsf{PKGen}(\mathsf{sk}_{j-1})\ (j\in[k(\lambda)]) \\
(\mathsf{cert},\mathsf{sk}^*,\mathsf{count}) \leftarrow \mathsf{Freeze}(\mathsf{sk}_{k(\lambda)})
\end{array}
\right] \geq 1 - \mathsf{negl}(\lambda).
\end{align}

\noindent\textbf{Decryption after freezing.} For any polynomial $k$ and any $(m_1,\dots,m_{k(\lambda)})\in\{0,1\}^{k(\lambda)}$,
\begin{align}
\Pr\left[
\begin{array}{@{}c@{}}
\forall i\in[k(\lambda)]\\
m_i\leftarrow\mathsf{Dec}(\mathsf{sk}^*,\mathsf{ct}_i)
\end{array}
:
\begin{array}{@{}l@{}}
\mathsf{crs} \leftarrow \mathsf{Setup}(1^\lambda) \\
(\mathsf{vk},\mathsf{sk}_0) \leftarrow \mathsf{KeyGen}(1^\lambda) \\
(\mathsf{pk}_j,\mathsf{sk}_j) \leftarrow \mathsf{PKGen}(\mathsf{sk}_{j-1})\ (j\in[k(\lambda)]) \\
(\mathsf{cert},\mathsf{sk}^*,\mathsf{count}) \leftarrow \mathsf{Freeze}(\mathsf{sk}_{k(\lambda)})\\
\mathsf{ct}_i\leftarrow\mathsf{Enc}(\mathsf{vk},\mathsf{pk}_i,m_i)\ (i\in[k(\lambda)])
\end{array}
\right] \geq 1 - \mathsf{negl}(\lambda),
\end{align}
where the $k(\lambda)$ executions of $\mathsf{Dec}$ take the classical part of $\mathsf{sk}^*$ as input, as in the correctness of \cref{def:QPKEcrs}.

\noindent\textbf{Index consistency.} Here we allow $\mathsf{Enc}$ to output a special symbol $\bot$, which means that $\mathsf{Enc}$ rejects the given public key. For any $\mathsf{crs}$, any $\mathsf{vk}$, any $m\in\{0,1\}$, and any quantum state $\rho$ given to $\mathsf{Enc}$ as the public key,
\begin{align}
\Pr\left[\mathsf{ct}\neq\bot\wedge\mathsf{Index}(\mathsf{vk},\mathsf{ct})=\bot : \mathsf{ct}\leftarrow\mathsf{Enc}(\mathsf{vk},\rho,m)\right]=0.
\label{eq:indexconsistency}
\end{align}
That is, $\mathsf{Enc}$ never outputs a ciphertext that is rejected by $\mathsf{Index}$.
\mor{I have added this new requirement, index consistency. Without it, $\mathsf{Enc}$ could output ciphertexts rejected by $\mathsf{Index}$, and the freeze would give no guarantee to the senders: a malicious receiver could keep handing out ``public keys'' after the freeze, and the honest senders would output ciphertexts that the receiver can decrypt but $\mathsf{Index}$ rejects, which is allowed by \cref{def:ctunique,def:freezebinding}.
The index consistency rules this out by requiring that every ciphertext output by $\mathsf{Enc}$, other than $\bot$, be accepted by $\mathsf{Index}$, so that \cref{lem:freeze_count} applies to the ciphertexts of the honest senders.
To achieve this definition, I have added an extra step in $\mathsf{Enc}'$: run $\mathsf{Index}$ on the ciphertext it has computed and output $\bot$ if $\mathsf{Index}$ rejects it.}
\end{definition}

\begin{definition}[Ciphertext uniqueness]
\label{def:ctunique}
We say that a QPKE scheme in the CRS model augmented with $\mathsf{Index}$ satisfies ciphertext uniqueness if for any QPT adversary $\mathcal{A}$, the probability that $\mathcal{A}$ wins the following experiment is $\le\mathsf{negl}(\lambda)$.
\begin{enumerate}
\item $\mathsf{crs}\leftarrow\mathsf{Setup}(1^\lambda)$.
\item $\mathcal{A}(1^\lambda,\mathsf{crs})$ outputs a verification key $\mathsf{vk}$ and two ciphertexts $\mathsf{ct}_1$ and $\mathsf{ct}_2$.
\item Run $\ell_i\leftarrow\mathsf{Index}(\mathsf{vk},\mathsf{ct}_i)$ for $i\in\{1,2\}$.
\item $\mathcal{A}$ wins if $\ell_1=\ell_2\neq\bot$ and $\mathsf{ct}_1\neq\mathsf{ct}_2$.
\end{enumerate}
\end{definition}

\begin{definition}[Freeze binding]
\label{def:freezebinding}
We say that a QPKE scheme in the CRS model augmented with $(\mathsf{Freeze},\mathsf{FreezeVer},\mathsf{Index})$ satisfies freeze binding 
if for any QPT adversary $\mathcal{A}$, the probability that $\mathcal{A}$ wins the following experiment is $\le\mathsf{negl}(\lambda)$.
\begin{enumerate}
\item $\mathsf{crs}\leftarrow\mathsf{Setup}(1^\lambda)$.
\item $\mathcal{A}(1^\lambda,\mathsf{crs})$ outputs a verification key $\mathsf{vk}$, a ciphertext $\mathsf{ct}$, a certificate $\mathsf{cert}$, and a count $\mathsf{count}$.
\item Run $b\leftarrow\mathsf{FreezeVer}(\mathsf{vk},\mathsf{cert},\mathsf{count})$ and $\ell\leftarrow\mathsf{Index}(\mathsf{vk},\mathsf{ct})$.
\item $\mathcal{A}$ wins if $b=\top$, $\ell\neq\bot$, and $\mathsf{count}<\ell$.
\end{enumerate}
\end{definition}

Ciphertext uniqueness and freeze binding together bound the number of ciphertexts accepted by $\mathsf{Index}$: once a certificate with a count $\mathsf{count}$ is accepted by $\mathsf{FreezeVer}$ under $\mathsf{vk}$, no QPT adversary, including the receiver itself, can produce more than $\mathsf{count}$ distinct ciphertexts accepted by $\mathsf{Index}$ under $\mathsf{vk}$, except with negligible probability.

\begin{lemma}
\label{lem:freeze_count}
Suppose that a QPKE scheme in the CRS model augmented with $(\mathsf{Freeze},\mathsf{FreezeVer},\mathsf{Index})$ satisfies ciphertext uniqueness and freeze binding.
Then for any QPT adversary $\mathcal{A}$, the probability that $\mathcal{A}$ wins the following experiment is $\le\mathsf{negl}(\lambda)$.
\begin{enumerate}
\item $\mathsf{crs}\leftarrow\mathsf{Setup}(1^\lambda)$.
\item $\mathcal{A}(1^\lambda,\mathsf{crs})$ outputs a verification key $\mathsf{vk}$, a certificate $\mathsf{cert}$, a count $\mathsf{count}$, and ciphertexts $\mathsf{ct}_1,\dots,\mathsf{ct}_N$.
\item $\mathcal{A}$ wins if $\top=\mathsf{FreezeVer}(\mathsf{vk},\mathsf{cert},\mathsf{count})$, $N>\mathsf{count}$, the ciphertexts $\mathsf{ct}_1,\dots,\mathsf{ct}_N$ are distinct, and $\mathsf{Index}(\mathsf{vk},\mathsf{ct}_i)\neq\bot$ for all $i\in[N]$.
\end{enumerate}
\end{lemma}

\begin{proof}
Let $\ell_i\coloneqq\mathsf{Index}(\mathsf{vk},\mathsf{ct}_i)$ for $i\in[N]$.
Whenever $\mathcal{A}$ wins, either $\ell_i>\mathsf{count}$ for some $i\in[N]$, or $\ell_i\le\mathsf{count}$ for all $i\in[N]$.
In the latter case, the $N$ indices $\ell_1,\dots,\ell_N$ take values in $\{1,\dots,\mathsf{count}\}$, and hence $\ell_i=\ell_j$ for some $i\neq j$ by the pigeonhole principle, because $N>\mathsf{count}$.
Let $\mathcal{B}_1$ be the QPT adversary that runs $\mathcal{A}$, computes $\ell_1,\dots,\ell_N$, and outputs $(\mathsf{vk},\mathsf{ct}_i,\mathsf{cert},\mathsf{count})$ for an $i$ with $\ell_i>\mathsf{count}$ if such $i$ exists.
Let $\mathcal{B}_2$ be the QPT adversary that runs $\mathcal{A}$, computes $\ell_1,\dots,\ell_N$, and outputs $(\mathsf{vk},\mathsf{ct}_i,\mathsf{ct}_j)$ for a pair $i\neq j$ with $\ell_i=\ell_j\neq\bot$ if such a pair exists.
$\mathcal{B}_1$ and $\mathcal{B}_2$ are QPT, because $N$ is bounded by a polynomial in $\lambda$ as $\mathcal{A}$ is QPT.
The probability that $\mathcal{A}$ wins is at most the sum of the probability that $\mathcal{B}_1$ wins the experiment of \cref{def:freezebinding} and the probability that $\mathcal{B}_2$ wins the experiment of \cref{def:ctunique}, both of which are negligible.
\end{proof}

\cref{lem:freeze_count} is stated for ciphertexts accepted by $\mathsf{Index}$.
Combined with the index consistency, it also bounds the ciphertexts generated by honest senders who are given public keys by a malicious receiver.

\begin{corollary}
\label{cor:freeze_count}
Suppose that a QPKE scheme in the CRS model augmented with $(\mathsf{Freeze},\mathsf{FreezeVer},\mathsf{Index})$ satisfies index consistency, ciphertext uniqueness, and freeze binding.
Then for any QPT adversary $\mathcal{A}$, the probability that $\mathcal{A}$ wins the following experiment is $\le\mathsf{negl}(\lambda)$.
\begin{enumerate}
\item $\mathsf{crs}\leftarrow\mathsf{Setup}(1^\lambda)$.
\item $\mathcal{A}(1^\lambda,\mathsf{crs})$ outputs a verification key $\mathsf{vk}$.
\item $\mathcal{A}$ can make encryption queries: $\mathcal{A}$ sends a quantum state $\rho$ and a message $m\in\{0,1\}$, and the challenger returns $\mathsf{ct}\leftarrow\mathsf{Enc}(\mathsf{vk},\rho,m)$. The queries model honest senders who are given public keys by $\mathcal{A}$.
\item $\mathcal{A}$ outputs a certificate $\mathsf{cert}$ and a count $\mathsf{count}$.
\item $\mathcal{A}$ wins if $\top=\mathsf{FreezeVer}(\mathsf{vk},\mathsf{cert},\mathsf{count})$ and the number of distinct ciphertexts other than $\bot$ returned by the challenger is larger than $\mathsf{count}$.
\end{enumerate}
\end{corollary}

\begin{proof}
Let $\mathcal{A}'$ be the QPT adversary that runs $\mathcal{A}$, answers the encryption queries by running $\mathsf{Enc}$ by itself, and outputs $\mathsf{vk}$, $\mathsf{cert}$, $\mathsf{count}$, and the distinct ciphertexts other than $\bot$ returned to $\mathcal{A}$.
By the index consistency, all these ciphertexts are accepted by $\mathsf{Index}$.
Hence $\mathcal{A}'$ wins the experiment of \cref{lem:freeze_count} whenever $\mathcal{A}$ wins, and the claim follows from \cref{lem:freeze_count}.
\end{proof}

\begin{remark}
The freeze does not prevent the receiver from handing out public keys after the freeze.
For example, in the construction of \cref{sec:freeze_construction}, a malicious receiver can sign the chain messages and the stop message first, publish the certificate with a count $\mathsf{count}$, and only then generate the public keys of the original scheme and combine them with the unused signing keys $\widetilde{\mathsf{sigk}}_1,\dots,\widetilde{\mathsf{sigk}}_{\mathsf{count}}$ to form $\mathsf{pk}'_1,\dots,\mathsf{pk}'_{\mathsf{count}}$ of \cref{eq:pk_freeze}.
What the freeze fixes is the number $\mathsf{count}$ of positions, and hence the number of distinct ciphertexts accepted by $\mathsf{Index}$ (\cref{lem:freeze_count,cor:freeze_count}).
Note also that the bound is on the number of distinct ciphertexts, and not on the number of executions of $\mathsf{Enc}$, the number of quantum states handed out as public keys, or the number of times the same ciphertext is reused.
\end{remark}

\subsection{Construction}
\label{sec:freeze_construction}

We use the following security of one-shot signatures, which is stronger than that of \cref{def:OSS}.

\begin{definition}[Strong security of one-shot signatures]
\label{def:strongOSS}
A one-shot signature scheme $(\mathsf{Setup},\mathsf{Gen},\mathsf{Sign},\mathsf{Ver})$ satisfies strong security if for any QPT adversary $\mathcal{A}$,
\begin{align}
\Pr\left[
\begin{array}{l}
\top\leftarrow\mathsf{Ver}(\mathsf{crs},\mathsf{vk},m_0,\sigma_0)\\
\wedge\,
\top\leftarrow\mathsf{Ver}(\mathsf{crs},\mathsf{vk},m_1,\sigma_1)\\
\wedge\,
(m_0,\sigma_0)\neq(m_1,\sigma_1)
\end{array}
:
\begin{array}{l}
\mathsf{crs}\leftarrow\mathsf{Setup}(1^\lambda)\\
(\mathsf{vk},m_0,\sigma_0,m_1,\sigma_1)\leftarrow\mathcal{A}(1^\lambda,\mathsf{crs})
\end{array}
\right]
\le\mathsf{negl}(\lambda).
\label{eq:strongOSS}
\end{align}
\end{definition}

As noted in \cref{rem:OSS}, the known constructions of one-shot signatures~\cite{C:ShmZha25,EPRINT:HuaVai25} satisfy strong security.

\begin{theorem}
\label{thm:freeze_construction}
If one-shot signatures with strong security and QPKE with quantum public keys exist, then there exists a QPKE scheme with quantum public keys 
in the CRS model that supports certified freeze (\cref{def:freeze}) and satisfies ciphertext uniqueness (\cref{def:ctunique}) and freeze binding (\cref{def:freezebinding}).
Moreover, if the original QPKE scheme satisfies everlasting security, then so does the constructed one.
\end{theorem}

Because strong security implies the security of \cref{def:OSS}, we obtain the following corollary from \cref{thm:OSS_to_QPKE,thm:freeze_construction}.

\begin{corollary}
\label{cor:freeze_construction}
If one-shot signatures with strong security exist, then there exists an everlasting-secure QPKE scheme with quantum public keys in the CRS model
that supports certified freeze and satisfies ciphertext uniqueness and freeze binding.
\end{corollary}

\begin{proof}[Proof of \cref{thm:freeze_construction}]
Let $(\mathsf{KeyGen},\mathsf{PKGen},\mathsf{Enc},\mathsf{Dec})$ be a QPKE scheme with quantum public keys, and let $(\mathsf{Setup},\mathsf{Gen},\mathsf{Sign},\mathsf{Ver})$ be a one-shot signature scheme with strong security.

We construct a QPKE scheme $(\mathsf{Setup}',\mathsf{KeyGen}',\mathsf{PKGen}',\mathsf{Enc}',\mathsf{Dec}')$ in the CRS model with $(\mathsf{Freeze},\mathsf{FreezeVer},\mathsf{Index})$ as follows.
\begin{itemize}
\item $\mathsf{Setup}'(1^\lambda)$: Run the setup algorithm of the one-shot signature scheme on input $1^\lambda$, and output its output $\mathsf{crs}$.
\item $\mathsf{KeyGen}'(1^\lambda)$: Run $(\mathsf{vk},\mathsf{sk}_0)\leftarrow\mathsf{KeyGen}(1^\lambda)$ and $(w_1,\mathsf{sigk}_1)\leftarrow\mathsf{Gen}(\mathsf{crs})$. Output $\mathsf{vk}'\coloneqq(\mathsf{vk},w_1)$ and $\mathsf{sk}'_0\coloneqq(\mathsf{sk}_0,(\,),\mathsf{sigk}_1)$, where $(\,)$ is the empty list of pairs.
\item $\mathsf{PKGen}'(\mathsf{sk}'_{t-1})$ (the $t$-th invocation): Parse $\mathsf{sk}'_{t-1}=(\mathsf{sk}_{t-1},(\mu_s,\tau_s)_{s\le t-1},\mathsf{sigk}_t)$.
\begin{enumerate}
\item Run $(\mathsf{pk}_t,\mathsf{sk}_t)\leftarrow\mathsf{PKGen}(\mathsf{sk}_{t-1})$.
\item Run $(v_t,\widetilde{\mathsf{sigk}}_t)\leftarrow\mathsf{Gen}(\mathsf{crs})$ and $(w_{t+1},\mathsf{sigk}_{t+1})\leftarrow\mathsf{Gen}(\mathsf{crs})$.
\item Set $\mu_t\coloneqq(t,v_t,w_{t+1})$ and run $\tau_t\leftarrow\mathsf{Sign}(\mathsf{crs},\mathsf{sigk}_t,\mu_t)$. This consumes $\mathsf{sigk}_t$.
Here and below, a tuple is regarded as a bit string from which all its components are uniquely recovered (for example, the concatenation of the components together with their lengths). In particular, two tuples are equal as bit strings if and only if all their components are equal.
\item Output
\begin{align}
\mathsf{pk}'_t\coloneqq\left(\mathsf{pk}_t,\widetilde{\mathsf{sigk}}_t,t,(\mu_s,\tau_s)_{s\le t}\right)
\label{eq:pk_freeze}
\end{align}
and $\mathsf{sk}'_t\coloneqq(\mathsf{sk}_t,(\mu_s,\tau_s)_{s\le t},\mathsf{sigk}_{t+1})$.
We write $\mathbf{Q}$ and $\mathbf{S}$ for the registers of $\mathsf{pk}'_t$ that contain $\mathsf{pk}_t$ and $\widetilde{\mathsf{sigk}}_t$, respectively.
\end{enumerate}
\item $\mathsf{Enc}'(\mathsf{vk}',\mathsf{pk}',m)$: Parse $\mathsf{vk}'=(\mathsf{vk},w_1)$.
\begin{enumerate}
\item Measure all the registers of $\mathsf{pk}'$ except for $\mathbf{Q}$ and $\mathbf{S}$ in the computational basis to obtain a bit string $x$. (For the honestly generated $\mathsf{pk}'_t$, $x=(t,(\mu_s,\tau_s)_{s\le t})$.)
\item Run $\mathsf{ct}\leftarrow\mathsf{Enc}(\mathsf{vk},\mathbf{Q},m)$, where the state of the register $\mathbf{Q}$ is used as the public key.
\item Run $\pi\leftarrow\mathsf{Sign}(\mathsf{crs},\mathbf{S},\mathsf{ct})$, where the state of the register $\mathbf{S}$ is used as the signing key.
\item Set
\begin{align}
\mathsf{ct}'\coloneqq(x,\mathsf{ct},\pi).
\label{eq:ct_freeze}
\end{align}
If $\mathsf{Index}(\mathsf{vk}',\mathsf{ct}')=\bot$, where $\mathsf{Index}$ is defined below, output $\bot$. Otherwise, output $\mathsf{ct}'$.
\end{enumerate}
\item $\mathsf{Dec}'(\mathsf{sk}'_t,\mathsf{ct}')$: Parse $\mathsf{sk}'_t=(\mathsf{sk}_t,(\mu_s,\tau_s)_{s\le t},\mathsf{sigk}_{t+1})$ and $\mathsf{ct}'=(x,\mathsf{ct},\pi)$. Run $m'\leftarrow\mathsf{Dec}(\mathsf{sk}_t,\mathsf{ct})$ and output $m'$. If the parsing of $\mathsf{ct}'$ fails, output $0$. The post-freeze secret key $\mathsf{sk}^*=(\mathsf{sk}_t,(\mu_s,\tau_s)_{s\le t})$ is used in the same way.
\item $\mathsf{Freeze}(\mathsf{sk}'_t)$: Parse $\mathsf{sk}'_t=(\mathsf{sk}_t,(\mu_s,\tau_s)_{s\le t},\mathsf{sigk}_{t+1})$. Run $\tau^*\leftarrow\mathsf{Sign}(\mathsf{crs},\mathsf{sigk}_{t+1},\bot)$,
where $\bot$ is a fixed bit string that is not the encoding of any tuple.
This consumes $\mathsf{sigk}_{t+1}$. Output $\mathsf{cert}\coloneqq((\mu_s,\tau_s)_{s\le t},\tau^*)$, $\mathsf{count}\coloneqq t$, and the post-freeze secret key $\mathsf{sk}^*\coloneqq(\mathsf{sk}_t,(\mu_s,\tau_s)_{s\le t})$.
\item $\mathsf{FreezeVer}(\mathsf{vk}',\mathsf{cert},\mathsf{count})$: Parse $\mathsf{vk}'=(\mathsf{vk},w_1)$ and $\mathsf{cert}=((\mu_s,\tau_s)_{s\le\mathsf{count}},\tau^*)$, and parse each $\mu_s$ as $(s,v_s,w_{s+1})$, i.e., the first component of $\mu_s$ has to be $s$. If the parsing fails, output $\bot$. Check $\top=\mathsf{Ver}(\mathsf{crs},w_s,\mu_s,\tau_s)$ for all $s\in\{1,\dots,\mathsf{count}\}$ and $\top=\mathsf{Ver}(\mathsf{crs},w_{\mathsf{count}+1},\bot,\tau^*)$. Output $\top$ if and only if all the checks pass.
\item $\mathsf{Index}(\mathsf{vk}',\mathsf{ct}')$: Parse $\mathsf{vk}'=(\mathsf{vk},w_1)$, $\mathsf{ct}'=(x,\mathsf{ct},\pi)$, and $x=(t,(\mu_s,\tau_s)_{s\le t})$ with $t\ge1$, and parse each $\mu_s$ as $(s,v_s,w_{s+1})$, i.e., the first component of $\mu_s$ has to be $s$. If the parsing fails, output $\bot$. Check $\top=\mathsf{Ver}(\mathsf{crs},w_s,\mu_s,\tau_s)$ for all $s\in\{1,\dots,t\}$ and $\top=\mathsf{Ver}(\mathsf{crs},v_t,\mathsf{ct},\pi)$. Output $t$ if all the checks pass, and output $\bot$ otherwise.
\end{itemize}
All algorithms are QPT, and $\mathsf{FreezeVer}$ and $\mathsf{Index}$ are classical and deterministic, because so is $\mathsf{Ver}$.
The classical part of $\mathsf{sk}'_t$ contains that of $\mathsf{sk}_t$, and hence $\mathsf{Dec}'$ acts only on the classical part of the secret key.
\paragraph{Correctness.}
Consider the experiment of the correctness, and write $k$ for $k(\lambda)$.
For each $i\in[k]$, $\mathsf{Enc}'(\mathsf{vk}',\mathsf{pk}'_i,m_i)$ computes $(x_i,\mathsf{ct}_i,\pi_i)$ with $x_i=(i,(\mu_s,\tau_s)_{s\le i})$ and $\mathsf{ct}_i\leftarrow\mathsf{Enc}(\mathsf{vk},\mathsf{pk}_i,m_i)$, and outputs it unless $\mathsf{Index}$ rejects it.
We first bound the probability that $\mathsf{Index}$ rejects $(x_i,\mathsf{ct}_i,\pi_i)$.
Since $\mu_s=(s,v_s,w_{s+1})$, the parsing in $\mathsf{Index}$ succeeds, and the checks of $\mathsf{Index}$ are the verifications of the $i+1$ pairs $(\mu_s,\tau_s)_{s\le i}$ and $(\mathsf{ct}_i,\pi_i)$, each of which is generated by the honest signing algorithm with an honestly generated signing key that has never signed before.
By the correctness of the one-shot signature scheme and the union bound, all the checks pass with probability $1-\mathsf{negl}(\lambda)$, in which case $\mathsf{Enc}'$ outputs $(x_i,\mathsf{ct}_i,\pi_i)$ and $\mathsf{Index}(\mathsf{vk}',(x_i,\mathsf{ct}_i,\pi_i))=i$.
Here we use the correctness for adversarially chosen messages, because the messages $\mu_s$ are correlated with $\mathsf{crs}$.
Hence, by the union bound over $i\in[k]$, $\mathsf{Enc}'$ outputs $\bot$ for some $i\in[k]$ with probability at most $k\cdot\mathsf{negl}(\lambda)=\mathsf{negl}(\lambda)$.
If $\mathsf{Enc}'$ outputs $(x_i,\mathsf{ct}_i,\pi_i)$, then $\mathsf{Dec}'(\mathsf{sk}'_k,(x_i,\mathsf{ct}_i,\pi_i))$ outputs the output of $\mathsf{Dec}(\mathsf{sk}_k,\mathsf{ct}_i)$.
Hence, if $m_i\leftarrow\mathsf{Dec}'(\mathsf{sk}'_k,\mathsf{ct}'_i)$ fails for some $i\in[k]$, then $\mathsf{Enc}'$ outputs $\bot$ for some $i\in[k]$, or $m_i\leftarrow\mathsf{Dec}(\mathsf{sk}_k,\mathsf{ct}_i)$ fails for some $i\in[k]$.
The former occurs with probability at most $\mathsf{negl}(\lambda)$ as shown above, and the latter occurs with probability at most $\mathsf{negl}(\lambda)$ by the correctness of the original scheme, because $(\mathsf{vk},(\mathsf{pk}_j,\mathsf{sk}_j)_{j\le k})$ and $(\mathsf{ct}_i)_{i\le k}$ are distributed as in the original scheme.
Hence the correctness follows.

\paragraph{Security.}
Let $\mathcal{A}'$ be an adversary against the constructed scheme.
We construct an adversary $\mathcal{A}$ against the original scheme as follows.
\begin{enumerate}
\item On receiving $\mathsf{vk}$ from the challenger, $\mathcal{A}$ runs $\mathsf{crs}\leftarrow\mathsf{Setup}'(1^\lambda)$ and $(w_1,\mathsf{sigk}_1)\leftarrow\mathsf{Gen}(\mathsf{crs})$, and sends $\mathsf{crs}$ and $\mathsf{vk}'\coloneqq(\mathsf{vk},w_1)$ to $\mathcal{A}'$.
\item When $\mathcal{A}'$ makes the $t$-th public-key query, $\mathcal{A}$ makes a public-key query to obtain $\mathsf{pk}_t$, runs the steps 2 and 3 of $\mathsf{PKGen}'$ by itself, and sends $\mathsf{pk}'_t$ of \cref{eq:pk_freeze} to $\mathcal{A}'$.
\item When $\mathcal{A}'$ sends the register $\mathbf{P}$ that is used as the public key for the challenge ciphertext (step 3 of the security game of \cref{def:QPKE}), $\mathcal{A}$ sends its subregister $\mathbf{Q}$ to the challenger, and receives $\mathsf{ct}$, where the subregisters $\mathbf{Q}$ and $\mathbf{S}$ of $\mathbf{P}$ are determined by the fixed layout of the public keys of the constructed scheme, as in $\mathsf{Enc}'$. $\mathcal{A}$ measures all the subregisters of $\mathbf{P}$ except for $\mathbf{Q}$ and $\mathbf{S}$ in the computational basis to obtain $x$, runs $\pi\leftarrow\mathsf{Sign}(\mathsf{crs},\mathbf{S},\mathsf{ct})$, and sets $\mathsf{ct}'\coloneqq(x,\mathsf{ct},\pi)$. $\mathcal{A}$ sends $\bot$ to $\mathcal{A}'$ if $\mathsf{Index}(\mathsf{vk}',\mathsf{ct}')=\bot$, and sends $\mathsf{ct}'$ to $\mathcal{A}'$ otherwise.
\item $\mathcal{A}$ outputs the output $b'$ of $\mathcal{A}'$.
\end{enumerate}
The measurement that yields $x$ and the execution of $\mathsf{Enc}$ on $\mathbf{Q}$ by the challenger act on different registers, and $\mathcal{A}$ computes $\mathsf{Index}$ on the classical data it holds.
Hence $\mathcal{A}$ perfectly simulates the security game of the constructed scheme for $\mathcal{A}'$ with the same challenge bit $b$.
Therefore the probability of $b'=b$ is the same in the two games.
If $\mathcal{A}'$ is QPT, so is $\mathcal{A}$, and the security of the constructed scheme follows from that of the original scheme.
If $\mathcal{A}'$ is an adversary of the form of \cref{def:everlastingQPKE}, so is $\mathcal{A}$, because all the operations added by $\mathcal{A}$ are QPT and are applied during the interaction with the challenger.
Hence the everlasting security of the constructed scheme follows from that of the original scheme.

\paragraph{Certified freeze.}
We first show the index validity. The experiment of the index validity is the same as that of the correctness except that only the $k(\lambda)$-th public key is used for encryption, and the argument in the proof of the correctness for $i=k(\lambda)$ applies verbatim: $\mathsf{Enc}'$ outputs $\mathsf{ct}'_{k(\lambda)}$ and $\mathsf{Index}(\mathsf{vk}',\mathsf{ct}'_{k(\lambda)})=k(\lambda)$ with probability $1-\mathsf{negl}(\lambda)$.

Next we show the freeze correctness. In the experiment of the freeze correctness, $\mathsf{count}=k(\lambda)$ by construction, the parsing in $\mathsf{FreezeVer}$ succeeds, and the checks of $\mathsf{FreezeVer}$ are the verifications of the $k(\lambda)+1$ pairs $(\mu_s,\tau_s)_{s\le k(\lambda)}$ and $(\bot,\tau^*)$, each of which is generated by the honest signing algorithm with an honestly generated signing key that has never signed before.
By the correctness of the one-shot signature scheme and the union bound, all the checks pass with probability $1-\mathsf{negl}(\lambda)$.

Next we show the decryption after freezing. $\mathsf{Freeze}$ acts only on $\mathsf{sigk}_{k(\lambda)+1}$, which is contained in neither $\mathsf{pk}'_i$ for $i\in[k(\lambda)]$ nor $\mathsf{sk}_{k(\lambda)}$, and hence the joint distribution of $(\mathsf{pk}'_i)_{i\le k(\lambda)}$ and $\mathsf{sk}_{k(\lambda)}$ is the same as in the experiment of the correctness.
Moreover, $\mathsf{sk}^*$ contains $\mathsf{sk}_{k(\lambda)}$, and $\mathsf{Dec}'$ uses only this component of the secret key.
Hence the executions of $\mathsf{Dec}'$ with $\mathsf{sk}^*$ are identical to those with $\mathsf{sk}'_{k(\lambda)}$, and the decryption after freezing follows from the correctness shown above.

Finally, the index consistency holds with probability $0$ in \cref{eq:indexconsistency} by construction: $\mathsf{Enc}'$ outputs a string other than $\bot$ only if $\mathsf{Index}$ accepts it, for any $\mathsf{crs}$, any $\mathsf{vk}'$, any $m$, and any quantum state given as the public key.

\paragraph{Ciphertext uniqueness.}
Let $\mathcal{A}$ be a QPT adversary that wins the experiment of \cref{def:ctunique} with probability $\epsilon$.
We construct a QPT adversary $\mathcal{B}$ against the strong security of the one-shot signature scheme as follows.
$\mathcal{B}(1^\lambda,\mathsf{crs})$ runs $(\mathsf{vk}',\mathsf{ct}'_1,\mathsf{ct}'_2)\leftarrow\mathcal{A}(1^\lambda,\mathsf{crs})$, and aborts unless $\mathsf{ct}'_1\neq\mathsf{ct}'_2$ and $\mathsf{Index}(\mathsf{vk}',\mathsf{ct}'_1)=\mathsf{Index}(\mathsf{vk}',\mathsf{ct}'_2)\neq\bot$.
Otherwise, let $t$ be the common output of $\mathsf{Index}$.
Then $\mathsf{vk}'$ parses as $(\mathsf{vk},w_1)$, and, for each $i\in\{1,2\}$, $\mathsf{ct}'_i$ parses as $((t,(\mu_{i,s},\tau_{i,s})_{s\le t}),\mathsf{ct}_i,\pi_i)$ with $\mu_{i,s}=(s,v_{i,s},w_{i,s+1})$. We set $w_{i,1}\coloneqq w_1$.
Exactly one of the following two cases occurs.
\begin{itemize}
\item Case 1: there exists $s\le t$ with $(\mu_{1,s},\tau_{1,s})\neq(\mu_{2,s},\tau_{2,s})$. Let $r$ be the smallest such index. Since $\mu_{1,s}=\mu_{2,s}$ for all $s<r$, we have $w_{1,r}=w_{2,r}$ (for $r=1$, both are $w_1$). $\mathcal{B}$ outputs $(w_{1,r},\mu_{1,r},\tau_{1,r},\mu_{2,r},\tau_{2,r})$.
\item Case 2: $(\mu_{1,s},\tau_{1,s})=(\mu_{2,s},\tau_{2,s})$ for all $s\le t$. Then $v_{1,t}=v_{2,t}$. 
Moreover, because $\mathsf{ct}'_1\neq\mathsf{ct}'_2$ as bit strings while their first components coincide, we have $(\mathsf{ct}_1,\pi_1)\neq(\mathsf{ct}_2,\pi_2)$.
$\mathcal{B}$ outputs $(v_{1,t},\mathsf{ct}_1,\pi_1,\mathsf{ct}_2,\pi_2)$.
\end{itemize}
In both cases, the two message-signature pairs output by $\mathcal{B}$ are distinct, and both pass the verification under the verification key output by $\mathcal{B}$, because $\mathsf{Index}$ has accepted both $\mathsf{ct}'_1$ and $\mathsf{ct}'_2$ and $\mathsf{Ver}$ is classical and deterministic.
Hence $\mathcal{B}$ wins the game of \cref{eq:strongOSS} with probability $\epsilon$, and therefore $\epsilon\le\mathsf{negl}(\lambda)$.
Note that the two pairs can differ only in the signatures, which is the reason why we need strong security.

\paragraph{Freeze binding.}
Let $\mathcal{A}$ be a QPT adversary that wins the experiment of \cref{def:freezebinding} with probability $\epsilon$.
We construct a QPT adversary $\mathcal{B}$ against the strong security of the one-shot signature scheme as follows.
$\mathcal{B}(1^\lambda,\mathsf{crs})$ runs $(\mathsf{vk}',\mathsf{ct}',\mathsf{cert},\mathsf{count})\leftarrow\mathcal{A}(1^\lambda,\mathsf{crs})$ and $t\leftarrow\mathsf{Index}(\mathsf{vk}',\mathsf{ct}')$, and aborts unless $\top=\mathsf{FreezeVer}(\mathsf{vk}',\mathsf{cert},\mathsf{count})$, $t\neq\bot$, and $\mathsf{count}<t$.
Otherwise, $\mathsf{vk}'$ parses as $(\mathsf{vk},w_1)$, $\mathsf{cert}$ parses as $((\mu_s,\tau_s)_{s\le\mathsf{count}},\tau^*)$ with $\mu_s=(s,v_s,w_{s+1})$, and $\mathsf{ct}'$ parses as $((t,(\mu'_s,\tau'_s)_{s\le t}),\mathsf{ct},\pi)$ with $\mu'_s=(s,v'_s,w'_{s+1})$. We set $w'_1\coloneqq w_1$.
Exactly one of the following two cases occurs.
\begin{itemize}
\item Case 1: there exists $s\le\mathsf{count}$ with $\mu'_s\neq\mu_s$. Let $r$ be the smallest such index. Since $\mu'_s=\mu_s$ for all $s<r$, we have $w'_r=w_r$ (for $r=1$, both are $w_1$). Then $\mathcal{B}$ outputs $(w_r,\mu_r,\tau_r,\mu'_r,\tau'_r)$.
\item Case 2: $\mu'_s=\mu_s$ for all $s\le\mathsf{count}$. Then $w'_{\mathsf{count}+1}=w_{\mathsf{count}+1}$ (for $\mathsf{count}=0$, both are $w_1$), and the position $\mathsf{count}+1$ exists in the chain of $\mathsf{ct}'$ because $\mathsf{count}<t$. Then $\mathcal{B}$ outputs $(w_{\mathsf{count}+1},\mu'_{\mathsf{count}+1},\tau'_{\mathsf{count}+1},\bot,\tau^*)$.
\end{itemize}
In both cases, the two messages output by $\mathcal{B}$ are distinct: in Case 1 by the definition of $r$, and in Case 2 because $\mu'_{\mathsf{count}+1}$ is an encoding of a tuple and $\bot$ is not.
Both signatures pass the verification under the verification key output by $\mathcal{B}$, because $\mathsf{FreezeVer}$ and $\mathsf{Index}$ have accepted and $\mathsf{Ver}$ is classical and deterministic.
Hence $\mathcal{B}$ wins the game of \cref{eq:strongOSS} with probability $\epsilon$, and therefore $\epsilon\le\mathsf{negl}(\lambda)$.
\end{proof}

\begin{remark}
\label{rem:freeze_security}
In the security game of \cref{def:QPKEcrs}, the adversary does not receive the certificate.
In practice, however, the receiver publishes the certificate, and hence the adversary can see it.
We therefore note that the security and the everlasting security of the constructed scheme hold even in the following extended game: at any point during the interaction with the challenger, the adversary can once instruct the challenger to run $\mathsf{Freeze}$ on the current secret key and receive $(\mathsf{cert},\mathsf{count})$, after which public-key queries are answered with $\bot$.
Indeed, the adversary $\mathcal{A}$ in the proof of \cref{thm:freeze_construction} holds all the signing keys of the one-shot signature scheme, including the unused $\mathsf{sigk}_{t+1}$ after $t$ public-key queries, and hence it can simulate $\mathsf{Freeze}$ by itself without any additional query to the challenger: it runs $\tau^*\leftarrow\mathsf{Sign}(\mathsf{crs},\mathsf{sigk}_{t+1},\bot)$ and returns $\mathsf{cert}=((\mu_s,\tau_s)_{s\le t},\tau^*)$ and $\mathsf{count}=t$ to $\mathcal{A}'$.
The rest of the simulation is unchanged, regardless of whether the freeze is requested before or after the challenge.
\end{remark}
%

\section{2QKE with Security for Alice}\label{sec:strong}

As is explained in \cref{sec:remarks_2QKE}, in the 2QKE constructed in the proof of \cref{thm:everlasting-ke},
Eve can learn Alice's key, and Eve can make Alice's key different from Bob's key.
In this section, we show that both issues are resolved if Bob additionally sends a short classical bit string,
which is a masked tag of an AMD code~\cite{EC:CDFPW08,DCC:CabPadSae02},
over the authenticated classical channel.

We consider the following variant of \cref{def:2QKE}:
Bob sends not only a quantum state but also a classical bit string to Alice, where the latter is sent over the authenticated classical channel,
and Alice outputs $\alpha\in\{0,1,\bot\}$, where $\bot$ means that Alice aborts.

\begin{definition}[Security for Alice]\label{def:strong}
A 2QKE satisfies security for Alice if for any QPT Eve who outputs a bit $e$,
\begin{align}
\Pr[\alpha\notin\{\beta,\bot\}]&\le\mathsf{negl}(\lambda), \label{eq:agreement}\\
\Pr[e=\alpha\wedge\alpha\neq\bot]&\le\frac{1}{2}\Pr[\alpha\neq\bot]+\mathsf{negl}(\lambda). \label{eq:secrecy}
\end{align}
Here, Eve can freely read the classical messages and can freely modify the quantum messages.
\end{definition}

We can actually show the following everlasting security:
\begin{definition}[Everlasting security for Alice]\label{def:strong_everlasting}
A 2QKE satisfies everlasting security for Alice if the following holds for any Eve of the following form:
during the execution of the protocol, Eve can freely read the classical messages and can freely modify the quantum messages, but all of her operations are QPT;
after the protocol terminates, Eve applies computationally-unbounded operations on all the registers that she keeps, and outputs a bit $e$.
The requirement is that \cref{eq:agreement} and \cref{eq:secrecy} hold.
\end{definition}

\paragraph{Construction.}
We construct a 2QKE that satisfies security for Alice (\cref{def:strong})
by modifying the 2QKE constructed in the proof of \cref{thm:everlasting-ke}.
All step numbers below refer to the latter.
We identify $\{0,1\}^\lambda$ with the finite field of order $2^\lambda$:
a bit string $(u_1,...,u_\lambda)$ is regarded as the polynomial $\sum_{i=1}^{\lambda}u_ix^{i-1}$ over $\mathbb{F}_2$
modulo a fixed irreducible polynomial of degree $\lambda$.
For $u,v\in\{0,1\}^\lambda$, $uv\in\{0,1\}^\lambda$ is the product in the field,
i.e., the product of the two polynomials modulo the irreducible polynomial.
(It is neither the inner product nor the bitwise product.)
The sum in the field is the bitwise XOR $u\oplus v\in\{0,1\}^\lambda$.
Let $n\coloneqq3\lambda$.
\begin{enumerate}
\item
Alice runs steps 1--6 independently $n$ times.
For each $j\in[n]$, let $\mathsf{vk}_j$, $a_j$, and $(\mathbf{A}_j,\mathbf{B}_j,\mathbf{C}_j)$ be the verification key, the bit $a$, and the registers of the $j$-th execution, respectively.
\item
Bob samples $u,v,w\gets\{0,1\}^\lambda$ and sets $k\coloneqq(u,v,w)\in\{0,1\}^n$.
For each $j\in[n]$, Bob performs the measurement $\{\Pi^{\mathsf{vk}_j},I-\Pi^{\mathsf{vk}_j}\}$ of step 7 on the registers of the $j$-th execution.
If the result is $I-\Pi^{\mathsf{vk}_j}$ for at least one $j$, Bob sends $\bot$ over the classical channel and the maximally-mixed state over the quantum channel.
Otherwise, Bob applies $Z^{k_j}$ on $\mathbf{A}_j$ for each $j\in[n]$, returns all the registers to Alice, and sends
\begin{align}
t\coloneqq uv\oplus w\in\{0,1\}^\lambda \label{eq:tag}
\end{align}
over the classical channel.
In both cases, Bob outputs the first bit of $u$ as $\beta$.
\item
If Alice receives $\bot$, she outputs $\alpha\coloneqq\bot$.
Otherwise, for each $j\in[n]$, Alice runs step 8 of the $j$-th execution, and let $k'_j$ be its output, i.e., the measurement result plus $a_j$.
She parses $k'=(u',v',w')$.
If $u'v'\oplus w'=t$, she outputs the first bit of $u'$ as $\alpha$. Otherwise, she outputs $\alpha\coloneqq\bot$.
\end{enumerate}

\begin{theorem}\label{thm:strong}
If one-time secure digital signatures with quantum signatures exist,
then the above 2QKE satisfies correctness, everlasting security (\cref{def:everlasting}), and 
everlasting security for Alice (\cref{def:strong_everlasting}).
\end{theorem}

\begin{remark}
In the above 2QKE, Eve cannot make Alice abort depending on Bob's key:
\begin{align}
\left|\Pr[\alpha\neq\bot\mid\beta=0]-\Pr[\alpha\neq\bot\mid\beta=1]\right|\le\mathsf{negl}(\lambda).
\end{align}
In fact, the proof of \cref{thm:strong} shows that, except for a negligible probability,
whether Alice aborts is determined by an event that is independent of Bob's key.
\end{remark}

The proof of \cref{thm:strong} is based on the following lemma, which shows that
the only thing that Eve can do on Alice's string $k'$ is to add a string that is independent of Bob's string $k$.

\begin{lemma}\label{lem:shift}
Let $\rho$ be the joint state of $k$, $\alpha$, and all the registers that Eve keeps when the protocol terminates.
Let $\rho'$ be the same state in the following modified execution, which is a hypothetical experiment used only in the proof:
Bob does not apply $Z^{k_j}$ on $\mathbf{A}_j$ for any $j\in[n]$.
Alice runs step 8 of the $n$ executions as in the real execution, and let $d\in\{0,1\}^n$ be the string of its outputs,
i.e., $d_j$ is Alice's measurement result of the $j$-th execution plus $a_j$.
Then $k'$ is defined as $k'\coloneqq d\oplus k$.
Then, for any QPT Eve,
\begin{align}
\|\rho-\rho'\|_1\le\mathsf{negl}(\lambda). \label{eq:shift}
\end{align}
\end{lemma}

\begin{proof}
As in the proof of \cref{thm:everlasting-ke}, we assume without loss of generality that
Eve applies a QPT unitary $V_1$ on the registers $(\mathbf{A}_j,\mathbf{B}_j,\mathbf{C}_j)_{j\in[n]}$ and her register $\mathbf{E}$ initialized to $|0...0\rangle$, and forwards $(\mathbf{A}_j,\mathbf{B}_j,\mathbf{C}_j)_{j\in[n]}$ to Bob.

Let $|\sigma_j(0)\rangle$, $|\sigma_j(1)\rangle$, and $P_j$ be the signatures and the Pauli of the $j$-th execution.
We can assume that Alice obtains $a_j$ in the following way:
instead of step 5, she generates
\begin{align}
\frac{1}{\sqrt{2}}\sum_{m\in\{0,1\}}|m\rangle_{\mathbf{R}_j}|m\rangle_{\mathbf{A}_j}|\sigma_j(m)\rangle_{\mathbf{B}_j}P_j|\sigma_j(1-m)\rangle_{\mathbf{C}_j},
\end{align}
keeps the register $\mathbf{R}_j$, and measures $\mathbf{R}_j$ in the Hadamard basis at the end of the protocol to obtain $a_j$.
In fact, $a_j$ is uniformly random, the state of the other registers is that of step 5, and Alice uses $a_j$ only at the end of the protocol.
Then, in the modified execution, adding $k_j$ to $d_j$ is equivalent to adding $k_j$ to $a_j$, which is equivalent to applying $Z^{k_j}$ on $\mathbf{R}_j$ before measuring it,
because $Z$ flips the result of the Hadamard-basis measurement.
Therefore the two executions are different only in the following point:
when Bob's measurement accepts, $Z^{k_j}$ is applied on $\mathbf{A}_j$ in the real execution, and on $\mathbf{R}_j$ in the modified execution.

Fix $k$, the verification keys, the signatures, and the Paulis.
For $m\in\{0,1\}^n$, define
\begin{align}
|\xi_m\rangle\coloneqq V_1\Big(\bigotimes_{j\in[n]}|m_j\rangle_{\mathbf{A}_j}|\sigma_j(m_j)\rangle_{\mathbf{B}_j}P_j|\sigma_j(1-m_j)\rangle_{\mathbf{C}_j}\otimes|0...0\rangle_{\mathbf{E}}\Big)|0...0\rangle_{\mathbf{B}'}, \label{eq:xi}
\end{align}
where $\mathbf{B}'$ is Bob's ancilla register,
so that the state before Bob's measurement is $2^{-n/2}\sum_{m\in\{0,1\}^n}|m\rangle_{\mathbf{R}}|\xi_m\rangle$ with $\mathbf{R}\coloneqq(\mathbf{R}_1,...,\mathbf{R}_n)$.
Let $\Pi\coloneqq\bigotimes_{j\in[n]}\Pi^{\mathsf{vk}_j}$, $Z^k\coloneqq\bigotimes_{j\in[n]}Z^{k_j}$, and $k\cdot m\coloneqq\bigoplus_{j\in[n]}k_jm_j$.
The unnormalized states after Bob's measurement accepts and $Z^k$ is applied on $\mathbf{A}\coloneqq(\mathbf{A}_1,...,\mathbf{A}_n)$ in the real execution and on $\mathbf{R}$ in the modified execution are
\begin{align}
|\Phi\rangle&\coloneqq\frac{1}{\sqrt{2^n}}\sum_{m\in\{0,1\}^n}|m\rangle_{\mathbf{R}}\otimes Z^k_{\mathbf{A}}\Pi|\xi_m\rangle,\\
|\Phi'\rangle&\coloneqq\frac{1}{\sqrt{2^n}}\sum_{m\in\{0,1\}^n}(-1)^{k\cdot m}|m\rangle_{\mathbf{R}}\otimes\Pi|\xi_m\rangle,
\end{align}
respectively.
All the subsequent operations are identical in the two executions.
Because $\||x\rangle\langle x|-|y\rangle\langle y|\|_1\le2\||x\rangle-|y\rangle\|$ for any vectors $|x\rangle$ and $|y\rangle$ of norm at most 1,
it therefore suffices to show that the expectation of $\||\Phi\rangle-|\Phi'\rangle\|$ over the fixed values is negligible.

We have
\begin{align}
\Pi|\xi_m\rangle=\sum_{m'\in\{0,1\}^n}\left(|m'\rangle\langle m'|_{\mathbf{A}}\otimes I\right)\Pi|\xi_m\rangle
\end{align}
and $Z^k_{\mathbf{A}}|m'\rangle_{\mathbf{A}}=(-1)^{k\cdot m'}|m'\rangle_{\mathbf{A}}$.
The terms with $m'=m$ of $|\Phi\rangle$ and $|\Phi'\rangle$ therefore cancel each other, and we obtain
\begin{align}
&|\Phi\rangle-|\Phi'\rangle\nonumber\\
&=\frac{1}{\sqrt{2^n}}\sum_{m}|m\rangle_{\mathbf{R}}\otimes\sum_{m'\neq m}\left((-1)^{k\cdot m'}-(-1)^{k\cdot m}\right)\left(|m'\rangle\langle m'|_{\mathbf{A}}\otimes I\right)\Pi|\xi_m\rangle.
\end{align}
The terms with different $(m,m')$ are orthogonal to each other, because they have different states on $(\mathbf{R},\mathbf{A})$,
and $|(-1)^{k\cdot m'}-(-1)^{k\cdot m}|\le2$.
Therefore
\begin{align}
\left\||\Phi\rangle-|\Phi'\rangle\right\|^2
\le\frac{4}{2^n}\sum_{m}\sum_{m'\neq m}\left\|\left(|m'\rangle\langle m'|_{\mathbf{A}}\otimes I\right)\Pi|\xi_m\rangle\right\|^2. \label{eq:shift-diff}
\end{align}
The right-hand side of \cref{eq:shift-diff} is four times the probability of the following event:
Alice samples $m\gets\{0,1\}^n$ and sends the state to which $V_1$ is applied in \cref{eq:xi},
Bob's measurement accepts, and the result of the computational-basis measurement of $\mathbf{A}$ is not $m$.
If the event occurs, there exists $j\in[n]$ such that the measurement $\{\Pi^{\mathsf{vk}_j},I-\Pi^{\mathsf{vk}_j}\}$ accepts and the result on $\mathbf{A}_j$ is $1-m_j$.
For each $j\in[n]$, the probability of the latter, averaged over the fixed values, is negligible from the one-time security of the digital signature scheme,
which is shown in the same way as the claim in the proof of \cref{thm:everlasting-ke}:
the adversary samples $k$ and $m$, outputs the message $m_j$, receives the signature $|\sigma_j(m_j)\rangle$,
generates the states of the other executions by itself, applies $V_1$, measures $\mathbf{A}_j$ in the computational basis, and outputs the state of $\mathbf{B}_j$ if the result is $1-m_j$.
(Because each $P_l$ is a uniformly random Pauli, the adversary can use a uniformly random computational-basis state instead of $P_l|\sigma_l(1-m_l)\rangle$.)
Hence, by the union bound, the expectation of the right-hand side of \cref{eq:shift-diff} is negligible,
and so is the expectation of $\||\Phi\rangle-|\Phi'\rangle\|$ by Jensen's inequality.
\end{proof}

\begin{proof}[Proof of \cref{thm:strong}]
Correctness follows from that of the 2QKE of \cref{thm:everlasting-ke} and the union bound:
if Eve does nothing, $k'=k$ with probability $1-\mathsf{negl}(\lambda)$, in which case $\alpha=\beta$.

For the security, we consider the modified execution of \cref{lem:shift}, where $k'=d\oplus k$.
We parse $d=(d_u,d_v,d_w)$.
Let $u_1$ be the first bit of $u$, i.e., $\beta=u_1$, and let $G$ be the event that Bob's measurement accepts and $d=0^n$.

We first show that, in the modified execution,
$(u,v)$ is independent of whether Bob's measurement accepts, $d$, and the state of the registers that Eve keeps.
The reason is as follows.
Whether Bob's measurement accepts does not depend on $k$, because Bob uses $k$ only after his measurement.
Because Bob does not apply $Z^{k_j}$, the only message that depends on $k$ is $t=uv\oplus w$ of \cref{eq:tag}, which Bob sends over the classical channel.
Hence $d$ and the state of the registers that Eve keeps depend on $k$ only through $t$,
and $t$ is independent of $(u,v)$ because $w$ is uniformly random.

Alice does not abort if and only if Bob's measurement accepts and $(u\oplus d_u)(v\oplus d_v)\oplus w\oplus d_w=uv\oplus w$, i.e.,
\begin{align}
ud_v\oplus vd_u\oplus d_ud_v\oplus d_w=0. \label{eq:check}
\end{align}
If $d=0^n$, \cref{eq:check} is satisfied.
If $d\neq0^n$, \cref{eq:check} is satisfied with probability at most $2^{-\lambda}$, which is shown as follows.
\begin{itemize}
\item
If $d_u=d_v=0^\lambda$, then $d_w\neq0^\lambda$, and \cref{eq:check} is $d_w=0^\lambda$. Hence it is not satisfied.
\item
If $d_v\neq0^\lambda$, \cref{eq:check} is $ud_v=vd_u\oplus d_ud_v\oplus d_w$.
Because $d_v$ has the inverse in the field, for each $v$ and $d$, there is exactly one $u$ that satisfies it.
Because $u$ is uniformly random and independent of $v$ and $d$, it is satisfied with probability $2^{-\lambda}$.
\item
If $d_u\neq0^\lambda$, \cref{eq:check} is satisfied with probability $2^{-\lambda}$ in the same way, where the roles of $u$ and $v$ are exchanged.
\end{itemize}
Therefore we obtain the following.
If $G$ occurs, Alice does not abort, and $\alpha=u_1$ because $u'=u$.
If $G$ does not occur, i.e., if Bob's measurement rejects or $d\neq0^n$, $\alpha=\bot$ except for probability $2^{-\lambda}$.

Hence $\alpha\in\{u_1,\bot\}$ except for probability $2^{-\lambda}$.
Because $\beta=u_1$, this means $\Pr[\alpha\notin\{\beta,\bot\}]\le2^{-\lambda}$.
Moreover,
\begin{align}
\Pr[e=\alpha\wedge\alpha\neq\bot]
&\le\Pr[e=u_1\wedge G]+2^{-\lambda}\\
&=\frac{1}{2}\Pr[G]+2^{-\lambda}
\le\frac{1}{2}\Pr[\alpha\neq\bot]+2^{-\lambda}.
\end{align}
Here, the first inequality is because $\alpha=u_1$ if $G$ occurs, and the probability that $G$ does not occur and $\alpha\neq\bot$ is at most $2^{-\lambda}$.
The equality is because $u_1$ is uniformly random and independent of $e$ and $G$
($e$ is obtained from the registers that Eve keeps, and $G$ is determined by whether Bob's measurement accepts and $d$), and therefore
\begin{align}
\Pr[e=u_1\wedge G]
&=\sum_{c\in\{0,1\}}\Pr[u_1=c]\Pr[e=c\wedge G]\\
&=\frac{1}{2}\sum_{c\in\{0,1\}}\Pr[e=c\wedge G]=\frac{1}{2}\Pr[G].
\end{align}
The last inequality is because $\alpha\neq\bot$ if $G$ occurs.
Moreover, because $\beta=u_1$ is uniformly random and independent of the state of the registers that Eve keeps,
$\Pr[e=\beta]=\frac{1}{2}$ even if $e$ is obtained by a computationally-unbounded operation on Eve's registers after the protocol terminates.

Finally, we show the theorem. In the real execution, we have
\begin{align}
\Pr[\alpha\notin\{\beta,\bot\}]&\le2^{-\lambda}+\mathsf{negl}(\lambda),\\
\Pr[e=\alpha\wedge\alpha\neq\bot]&\le\frac{1}{2}\Pr[\alpha\neq\bot]+\mathsf{negl}(\lambda),\\
\Pr[e=\beta]&\le\frac{1}{2}+\mathsf{negl}(\lambda),
\end{align}
which are \cref{eq:agreement}, \cref{eq:secrecy}, and the everlasting security, respectively.
The reason is as follows.
We have shown above that the same three inequalities hold in the modified execution without $\mathsf{negl}(\lambda)$.
From \cref{eq:shift}, the probability of each of the events
$\alpha\notin\{\beta,\bot\}$, $e=\alpha\wedge\alpha\neq\bot$, $\alpha\neq\bot$, and $e=\beta$
in the real execution is different from that in the modified execution only negligibly.
Here, we can use \cref{eq:shift} because each event is determined by $k$, $\alpha$, and the registers that Eve keeps:
$\beta$ is the first bit of $u$, which is a part of $k$, and $e$ is obtained from the registers that Eve keeps.
\end{proof}

\fi

\ifnum\anonymous=1
\else
\vspace{1cm}
\textbf{Acknowledgements.}
TM is supported by
JST CREST JPMJCR23I3,
JST Moonshot R\verb|&|D JPMJMS2061-5-1-1, 
JST FOREST, 
MEXT QLEAP, 
the Grant-in Aid for Transformative Research Areas (A) 21H05183,
and 
the Grant-in-Aid for Scientific Research (A) No.22H00522.
\fi

\textbf{AI Disclosure.}
We used 
Claude Fable 5.1 and ChatGPT-6 Astra to assist with
writing proofs and Introduction.
The authors verified all texts, and retain full responsibility for all content of the paper.

\ifnum\submission=0
\bibliographystyle{alpha} 
\else
\bibliographystyle{splncs04}
\fi
\bibliography{abbrev3,crypto,reference,text}

\appendix
\renewcommand{\theHsection}{\Alph{section}}
\crefalias{section}{appendix}
\section{Proof of \cref{lem:qots-from-evsp}}
\label{app:relations}

\begin{proof}
We first show item 1.
Let $(\mathsf{Mint},\mathsf{Vrfy})$ be a public-key quantum money mini-scheme.
We construct an EV-StatePuzz $(\mathsf{Samp},\mathsf{Ver})$ as follows.
\begin{itemize}
\item $\mathsf{Samp}(1^\lambda)$: Run $(s,\rho_s)\leftarrow\mathsf{Mint}(1^\lambda)$, and output the puzzle $s$ and the answer $\psi_s\coloneqq\rho_s$.
\item $\mathsf{Ver}(1^\lambda,s,\rho)$: Run $\mathsf{Vrfy}(1^\lambda,s,\rho)$ and output its output.
\end{itemize}

\paragraph{Correctness.}
The correctness experiment of $(\mathsf{Samp},\mathsf{Ver})$ is identical to that of $(\mathsf{Mint},\mathsf{Vrfy})$, and hence
\begin{align}
&\Pr[\top\gets\mathsf{Ver}(1^\lambda,s,\psi_s):(s,\psi_s)\gets\mathsf{Samp}(1^\lambda)]\\
&=\Pr[\top\gets\mathsf{Vrfy}(1^\lambda,s,\rho_s):(s,\rho_s)\gets\mathsf{Mint}(1^\lambda)]
\geq1-\mathsf{negl}(\lambda).
\end{align}

\paragraph{Security.}
Assume there exists a QPT $\mathcal{A}$ and a polynomial $p$ such that
\begin{align}
\Pr[\top\gets\mathsf{Ver}(1^\lambda,s,\mathcal{A}(1^\lambda,s)):(s,\psi_s)\gets\mathsf{Samp}(1^\lambda)]
\ge\frac{1}{p(\secp)}
\end{align}
for infinitely-many $\secp$.
We construct a QPT adversary $\mathcal{B}$ against the security of the mini-scheme as follows.
$\mathcal{B}(1^\lambda,s,\rho_s)$ runs $\rho\leftarrow\mathcal{A}(1^\lambda,s)$, and outputs the state $\sigma$ over two registers whose first register contains the received money state $\rho_s$ and whose second register contains $\rho$.
Because $\mathsf{Samp}=\mathsf{Mint}$, the input $s$ to $\mathcal{A}$ is distributed identically to that in the security experiment of the EV-StatePuzz.
In the execution of $\mathsf{Vrfy}_2(1^\lambda,s,\sigma)$, the two invocations of $\mathsf{Vrfy}(1^\lambda,s,\cdot)$ act on the two disjoint registers of $\sigma$, and therefore, by the union bound,
\begin{align}
\Pr\left[\top\leftarrow\mathsf{Vrfy}_2(1^\lambda,s,\sigma)\right]
&\geq1-\Pr[\bot\gets\mathsf{Vrfy}(1^\lambda,s,\rho_s)]-\Pr[\bot\gets\mathsf{Vrfy}(1^\lambda,s,\rho)]\\
&\geq\frac{1}{p(\lambda)}-\mathsf{negl}(\lambda),
\end{align}
where we used the correctness of the mini-scheme for the first register, which is untouched by $\mathcal{A}$, and $\Pr[\top\gets\mathsf{Vrfy}(1^\lambda,s,\rho)]\ge\frac{1}{p(\lambda)}$ for the second register.
This breaks the security of the money scheme.
This completes the proof of item 1.

\medskip
Next we show item 2.
Let $(\mathsf{Samp},\mathsf{PVer})$ be an EV-StatePuzz, where we write $\mathsf{PVer}$ for its verification algorithm in order to distinguish it from the verification algorithm of the signature scheme. 
We construct a one-time digital signature scheme $(\mathsf{KeyGen},\mathsf{Sign},\mathsf{Ver})$ as follows.
\begin{itemize}
    \item $\mathsf{KeyGen}(1^\lambda)$: For each $b\in\{0,1\}$, run $(s_b,\psi_b)\leftarrow\mathsf{Samp}(1^\lambda)$.
    Output $\mathsf{vk}\coloneqq(s_0,s_1)$ and $\mathsf{sigk}\coloneqq(\mathsf{sigk}_0,\mathsf{sigk}_1)$, where $\mathsf{sigk}_b$ is the register containing $\psi_b$.
    \item $\mathsf{Sign}(\mathsf{sigk},m)$: Output the register $\mathsf{sigk}_m$, whose state is $\sigma\coloneqq\psi_m$. 
    \item $\mathsf{Ver}(\mathsf{vk},m,\sigma)$: Run $\mathsf{PVer}(1^\lambda,s_m,\sigma)$ and output its output.
\end{itemize}

\paragraph{Correctness.}
An honestly generated signature on $m$ is $\psi_m$, which is the answer of the puzzle $s_m$, and $(s_m,\psi_m)$ is distributed as an output of $\mathsf{Samp}(1^\lambda)$.
Thus the correctness immediately follows from the correctness of the EV-StatePuzz.

\paragraph{One-time security.}
Assume that there exists a QPT $\mathcal{A}=(\mathcal{A}_1,\mathcal{A}_2)$ and a polynomial $p$
that breaks the one-time security with probability $1/p(\lambda)$ for infinitely many $\secp$.
We construct a QPT adversary $\mathcal{B}$ against the security of the EV-StatePuzz as follows.

$\mathcal{B}(1^\lambda,s^*)$:
\begin{enumerate}
    \item Sample $b^*\leftarrow\{0,1\}$ uniformly at random.
    \item Set $s_{b^*}\coloneqq s^*$ and run $(s_{1-b^*},\psi_{1-b^*})\leftarrow\mathsf{Samp}(1^\lambda)$.
    \item Set $\mathsf{vk}\coloneqq(s_0,s_1)$ and run $(m,\mathsf{st})\leftarrow\mathcal{A}_1(\mathsf{vk})$.
    \item If $m=b^*$, abort. Otherwise, set $\sigma\coloneqq\psi_{1-b^*}$ and run $\sigma'\leftarrow\mathcal{A}_2(\sigma,\mathsf{st})$.
    \item Output $\sigma'$.
\end{enumerate}
Note that $\mathcal{B}$ never uses the answer of the puzzle $s^*$: the signing query is answered only when $m=1-b^*$, in which case the signature is $\psi_{1-b^*}$, which $\mathcal{B}$ sampled by itself.

Since $s^*$ is distributed identically to the puzzle part of a fresh output of $\mathsf{Samp}(1^\lambda)$, and since the answer $\psi_{b^*}$ is discarded in the real experiment when $m=1-b^*$, the experiment that $\mathcal{B}$ simulates for $\mathcal{A}$ conditioned on not aborting is distributed identically to the real one-time security experiment.
Moreover, $b^*$ is uniformly random and independent of the entire execution of $\mathcal{A}$, and in particular of $m$.
Let $W$ be the event that $\mathcal{A}$ wins, i.e., $\mathsf{Ver}(\mathsf{vk},1-m,\sigma')$ outputs $\top$, so that $\Pr[W]=1/p(\lambda)$.
Conditioned on $b^*=1-m$, which occurs with probability $1/2$ independently of $W$, $\mathcal{B}$ does not abort, and $\mathcal{A}$ wins if and only if $\mathsf{PVer}(1^\lambda,s_{1-m},\sigma')=\mathsf{PVer}(1^\lambda,s^*,\sigma')$ outputs $\top$.
Therefore,
\begin{align}
\Pr\left[\top\leftarrow\mathsf{PVer}(1^\lambda,s^*,\rho) :
\begin{array}{l}
(s^*,\psi_{s^*})\leftarrow\mathsf{Samp}(1^\lambda)\\
\rho\leftarrow\mathcal{B}(1^\lambda,s^*)
\end{array}
\right]\geq\frac{1}{2p(\secp)}
\end{align}
for infinitely many $\secp$,
which breaks the security of the EV-StatePuzz. 
This completes the proof of item 2.
\end{proof}
\ifnum\submission=0
\else

\fi

\clearpage


\end{document}